\documentclass[prx,reprint,twocolumn,showpacs,superscriptaddress]{revtex4-2}
\usepackage{graphicx}
\usepackage{pstricks}
\usepackage{bbm}
\usepackage{float}
\usepackage{tikz}

\usetikzlibrary{calc}
\usetikzlibrary{shapes} 
\usetikzlibrary{
	trees,shadows,positioning,arrows,chains,
	decorations.pathreplacing,
	decorations.pathmorphing,
	decorations.shapes,
	decorations.text,
	shapes.geometric,
	shapes.symbols,
	matrix,
	math,
	patterns,
	intersections,
	fit}
\usetikzlibrary{arrows.meta}
\usetikzlibrary{decorations.markings}
\usepackage{amsmath,amssymb}
\usepackage{wasysym}
\usepackage{array, romannum}
\usepackage{caption}
\usepackage{textcomp,siunitx}
\DeclareSIUnit\gauss{G}
\usepackage[
	pdftex,
	colorlinks=true,
  	urlcolor=blue,
  	linkcolor=blue,
  	citecolor=blue
]{hyperref}
\usepackage{color}
\usepackage[T1]{fontenc}
\usepackage{currvita}

\newcommand{\ket}[1]{\left\lvert #1 \right\rangle}%
\newcommand{\sups}[2]{#1^{(#2)}}
\renewcommand{\vec}[1]{{\boldsymbol{#1}}}
\newcommand{\ga}{\gamma}
\newcommand{\ceq}{\overset{c}{=}}

\tikzset{global scale/.style={
		scale=#1,
		every node/.append style={scale=#1}
	}
}
\begin{document}

\title{Higher-dimensional Jordan-Wigner Transformation and Auxiliary Majorana Fermions}

\author{Kangle Li}
\affiliation{Department of Physics, Hong Kong University of Science and Technology, Clear Water Bay, Hong Kong, China}

\author{Hoi Chun Po}
\affiliation{Department of Physics, Hong Kong University of Science and Technology, Clear Water Bay, Hong Kong, China}
\date{\today}

\begin{abstract}
We discuss a scheme for performing Jordan-Wigner transformation for various lattice fermion systems in two and three dimensions which keeps internal and spatial symmetries manifest. 
The correspondence between fermionic and bosonic operators is established with the help of auxiliary Majorana fermions. The current construction is applicable to general lattices with even coordination number and an arbitrary number of fermion flavors.
The approach is demonstrated on the single-orbital square, triangular and cubic lattices for spin-$1/2$ fermions. We also discuss the relation to some quantum spin liquid models.
\end{abstract}

\maketitle

\section{INTRODUCTION}\label{sec_1}
Bosonization has long been an important technique for describing and understanding many-body quantum systems. For instance, it allows one to directly apply tools developed for bosonic systems, like tensor network methods, to fermionic problems.
Generally speaking, the fermionic statistics is implemented on the bosonic side using non-local string operators, as is well-known in the classic Jordan-Wigner transformation for one-dimensional systems \cite{JW}. Generalization of the transformation to higher dimensions using lattice gauge fields have also been proposed
 \cite{Wosiek:1981mn, cmp/1103942539, PhysRevLett.63.322, PhysRevB.43.3786, WENG200067, PhysRevLett.95.176407, Verstraete_2005, PhysRevB.75.144401, PhysRevLett.98.087204, PhysRevB.76.193101, Chen_2008, Cobanera2011, PhysRevB.86.085415, PhysRevB.98.075119, CHEN2018234, PhysRevB.100.245127,  PhysRevResearch.2.033527, PhysRevD.102.114502, Clifford}. Such exact bosonization mappings are also important form the quantum simulation perspective, for they address the possibility of simulating a quantum many-body fermionic systems using a bosonic quantum computer \cite{BRAVYI2002210, Seeley2012, setia2018bravyi, setia2019superfast, PhysRevB.104.035118, derby2021compact}. Symmetries, however, may not remain manifest under the transformation. For instance, while Refs.\ \onlinecite{CHEN2018234, PhysRevB.100.245127,  PhysRevResearch.2.033527, PhysRevD.102.114502, Clifford} discussed schemes for bosonizing spinless fermions on general lattices, it is unclear how they can be extended to cover spin-$1/2$ fermions, like the electrons, with spin-rotation invaraince.

In a recent work \cite{po2021symmetric}, we propose an approach for performing higher-dimensional Jordan-Wigner transformation while keeping all symmetries manifest.
The symmetry transformation on the bosonic side can all be traced down to that of a collection of operators denoted by $ \Lambda^{\alpha \beta}$, which could be viewed as the bosonic analog (up to a Jordan-Wigner string) of the Majorana fermions defining the physical fermionic Hilbert space. The approach in Ref.\ \onlinecite{po2021symmetric}, however, is limited to four-coordinated lattices like the 2D square and 3D diamond lattices with a single orbital per site.
In this work, we generalize our construction to lattices with even coordination number and an arbitrary number of orbitals.
Our construction follows from decomposing $\Lambda^{\alpha \beta} = i \eta^{\alpha} \chi^\beta$ using auxiliary Majorana fermions $\eta$ and $\chi$. Here, the $\eta$ fermions carry the physical quantum numbers and their bilinears generate internal symmetries. In contrast, the $\chi$ fermions encode the directional dependence in the bosonization scheme and their bilinears generate transformations which keep spatial symmetries manifest. 
For simplicity, we will refer to this as the $\eta\chi$ approach henceforth. 

The $\eta \chi$ approach is inspired by earlier works providing lattice bosonization recipes through the introduction of Majorana fermions \cite{PhysRevLett.95.176407, Verstraete_2005, setia2018bravyi}. It is also closely related to the approach developed in Refs.\ \onlinecite{Wosiek:1981mn, cmp/1103942539, PhysRevD.102.114502, Clifford}, dubbed the ``$\Gamma$ model,'' given that the $\eta$ and $ \chi$ Majorana fermions provide a natural representation of the Clifford algebra. 
In the following, we build on these earlier results and discuss how symmetries can remain manifest after bosonization. In addition, our approach allows for an arbitrary number of complex fermionic modes on each site and incorporates naturally any possible internal symmetries among the fermion flavos, although we will limit the majority of our discussion to lattices with even coordination numbers.

The paper is organized as follows: in Sec. \ref{general_procedures}, we first discuss the general procedures of our bosonization scheme. We define the operators invoked and discuss the fermion-boson mapping. After that we study examples. As a warm up, we revisit the bosonization problem on the square lattice in Sec. \ref{sec_2} within the $\eta\chi$ formalism. We review the constraints defining the fermionic states within the bosonic Hilbert space, and also discuss how the spatial symmetries are implemented. In Sec. \ref{sec_3} we provide a parallel discussion for the triangular lattice, which is six-coordinated and goes beyond the approach developed in Ref.\ \onlinecite{po2021symmetric}. For a single-orbital triangular lattice model with spin-$1/2$ fermions, $\Lambda^{ij}$ will be a $4\times6$ operator-valued matrix, i.e., we have four $\eta$'s and six $\chi$'s. 
The geometric frustration in the triangular lattice could lead to a spin liquid state at half-filling, and it is an interesting problem to explore the phases that emerge upon doping the system with electrons or holes \cite{PhysRevX.10.021042}.
In Sec. \ref{sec_4}, we further apply the $\eta\chi$ formalism to the 3D cubic lattice, which is also six-coordinated. The operator constraints defining the fermionic subspace is more complicated in three dimensions, and we provide a strategy for finding independent constraints. We also analyze in details how symmetries are implemented on the bosonized model, and discuss how fermion-odd operators can also be identified on the bosonic side \cite{po2021symmetric}. We conclude in Sec. \ref{sec_5} with a discussion on how the $\eta\chi$ formalism might be generalized to odd-coordinated systems, and elaborate on the relations of our approach with the design of exactly solved quantum spin liquid models \cite{KITAEV20062,PhysRevB.79.075124, PhysRevB.85.155119}.

\section{General procedures}\label{general_procedures}
Let us begin by introducing the general bosonization procedure in the $\eta\chi$ formalism. Given an arbitrary lattice fermionic system, which can be viewed as a connected graph consisting of some vertices and edges, we first consider the operators localized to each site $\vec r$. Suppose we have $m$ complex fermionc modes, $f^1_{\vec r},f^2_{\vec r},\cdots,f^m_{\vec r}$. We can represent them by $2m$ Majorana operators $\gamma^k_{\vec r},k=1,2,\cdots,2m$ through
 \begin{equation}
 f^i_{\vec r}=\frac{1}{2}(\ga^{2i-1}_{\vec r}-i\ga^{2i}_{\vec r});~~~ f^{i\dagger}_{\vec r}=\frac{1}{2}(\ga^{2i-1}_{\vec r} + i\ga^{2i}_{\vec r}),
 \end{equation}
 for $1\leq i\leq m$.
These Majorana fermions obey anticommutation relations
\begin{equation}
\begin{split}
	\{\gamma^{i}_{\vec{r}},\gamma^{j}_\vec{r'} \}=2\delta^{ij}\delta_{\vec{r}\vec{r'}}.
\end{split}
\end{equation}
All operators in fermions $f^i$ can be rewritten in terms of $\ga^i$. In particular, terms in the Hamiltonian can always be written as sum and product of the bilinears $i\gamma_\vec{r}^i\gamma_\vec{r'}^j$. 

The $\eta\chi$ formalism encodes the fermionic problem in a bosonic Hilbert space as follows. To each site, we attach Majorana operators $\eta^i_{\vec r},i\in\{1,2,\cdots 2m\}$, and $\chi^i_{\vec r},i\in\{1,2, n_\vec{r}\}$ where $n_\vec{r}$ is the coordination number of the site (for the examples we study in the following, $n_\vec{r}$ is a global constant independent of $\vec r$). We assume $n_\vec{r}$ is even for all $\vec{r}$. Then we define operators
\begin{equation}\label{parton_cons}
\begin{split}
\Theta^{ij}_{\vec{r}}=i\eta^i_{\vec{r}}\eta^j_{\vec{r}};\quad \Lambda^{ij}_{\vec{r}}=i\eta^i_{\vec{r}}\chi^j_{\vec{r}};\quad
 \Phi^{ij}_{\vec{r}}\equiv -i \chi^i_{\vec{r}}\chi^j_{\vec{r}}.
\end{split}
\end{equation}
From these expressions one can readily check that $\Theta, \Lambda$ and $\Phi$ satisfy the relations in Ref. \onlinecite{po2021symmetric}, 
\begin{equation}\label{mapping}
\Lambda^{ij}_{\vec{r}}=-i\Theta^{ij}_{\vec{r}}\Lambda^{jj}_{\vec{r}};\quad \Phi^{ij}_{\vec{r}}=-\Lambda^{ki}_{\vec{r}}\Theta^{kl}_{\vec{r}}\Lambda^{lj}_{\vec{r}},
\end{equation}
where there is no summation on the repeated indices. 
Both $\Theta^{ij}_{\vec{r}}$ and $\Lambda^{ij}_{\vec{r}}$ are Hermitian, meanwhile both $\Theta^{ij}_{\vec{r}} and \Phi^{ij}_{\vec{r}}$ are anti-symmetric in $i,j$.
Note that although $\Lambda^{ij}$ is a $4\times 4$ operator-valued matrix in Ref.\ \onlinecite{po2021symmetric}, in the current approach it can be a general even-by-even rectangular matrix.

Fermion bilinears in the physical problem can be mapped to the bosonic operators $\Theta$ and $\Lambda$ through
\begin{equation}\label{ga_map}
\begin{split}
i\gamma^i_{\vec{r}} \gamma^j_{\vec{r}} &\rightarrow \Theta^{ij}_{\vec{r}}; \\
i\gamma^i_{\vec{r}} \gamma^j_{\vec{r'}} &\rightarrow \Lambda^{ix}_{\vec{r}} \Lambda^{jy}_{\vec{r'}},
\end{split}
\end{equation}
where $\vec{r}$ and $\vec{r'}$ are the ending  and starting sites of a given arrow (an oriented edge) in the lattice
\begin{tikzpicture}
\draw[dashed] (0,0)--(0.5,0);
\draw[->, dashed] (1,0)--(0.5,0);
\node[left] at(0,0){$\vec{r}$};
\node[right] at(1,0){$\vec{r'}$};
\end{tikzpicture}. 
Though not necessary, we would often take the edges to be between nearest neighbors. Here, $x,y$ are numbers labeling the Majorana $\chi^{x,y}$ at two ends of the edge. We also demand $\Theta^{ij}$ and $\Lambda^{ij}$ to satisfy the same commutation and anticommutation relations as the femrion bilinears they represent; the two operators  $\Theta^{ij}$ and $\Lambda^{lm}$ anticommute if and only if $l$ equals to either $i$ or $j$; otherwise they commute. 
Similarly, the two operators  $\Phi^{ij}$ and $\Lambda^{lm}$ anticommute  if $m$ equals to $i$ or $j$, and commute otherwise. The two operators $\Theta^{ij}$ and $\Phi^{lm}$ always commute.  For applications we will also discuss qubit representations of $\Theta^{ij}$ and $\Lambda^{lm}$ in following sections.

In this construction, $\eta^i$'s correspond to physical degrees of freedom, while $\chi^i$'s are auxiliary Majorana fermions connecting different sites. The number of $\eta^i$'s and the number of $\chi^i$'s are not necessarily the same. This is different from the formalism in Ref. \onlinecite{po2021symmetric}, which relied on an exceptional isomorphism of the group ${\rm Spin}(4)$. The internal symmetries like femion parity, time reversal, flavor symmetries, are characterized by transformations of $\eta^i$'s. Spatial symmetries, like translation, reflection, and rotations, are characterized by transformations of $\chi^i$'s. 

To be more explicit, let us take a closer look at the transformations of $\Lambda^{ij}=i\eta^i\chi^j$. The bilinear operators $\theta^{ij}=\frac{i}{2}\eta^i\eta^j$ form a set of generators of $\mathfrak{so}(2m)$ algebra
\begin{equation}\label{comm}
[\theta^{ij},\theta^{kl}]=i\left( \delta^{il}\theta^{jk}+\delta^{jk}\theta^{il}-\delta^{ik}\theta^{jl}-\delta^{jl}\theta^{ik} \right).
\end{equation}
The exponentiation of elements in the algebra form Spin($2m$) which is the double cover of SO($2m$). A similar structure can be introduced for $\phi^{ij}_\vec{r}=-\frac{i}{2} \chi^{i}\chi^{j}$ which form a set of generators of $\mathfrak{so}(2n)$ algebra. Furthermore, we have
\begin{equation}\label{eta_vec}
    [\theta^{ij},\eta^k]=i(\delta^{jk}\eta^i-\delta^{ik}\eta^{j}), \quad i\neq j.
\end{equation}
 An element in Spin($2m$) can be written as $U(A)=e^{-\sum_{ij}A_{ij}\theta^{ij}}$, where $A$ is a $2m\times 2m$ anti-symmetric matrix. Use Eq. \ref{eta_vec}, we see
\begin{equation}
    U(A)\eta^kU(A)^{-1}=\sum_{k'}\left(e^{-2A}\right)_{kk'}\eta^{k'}.
\end{equation}
The matrix $e^{-2A}$ is an element in SO($2m$), meaning $\eta^k$ transforms as an SO($2m$) vector. Similarly we define $V(A)=e^{-\sum{ij}A_{ij}\phi^{ij}}$. $\chi^i$ transforms as an SO($2n$) vector. Combining these properties, $\Lambda^{ij}=i\eta^i\chi^j$ will transform as an SO($2m$) vector in its first index and as an SO($2n$) vector in its second index independently,
\begin{equation}
    \begin{split}
        U(A)\Lambda^{ij}U(A)^{-1} & =\sum_{i'}\left(e^{-2A}\right)_{ii'}\Lambda^{i'j}.\\
        V(A)\Lambda^{ij}V(A)^{-1} & =\sum_{j'}\Lambda^{ij'}\left(e^{-2A}\right)_{j'j}.\\
    \end{split}
\end{equation}
So to find transformations among different $\chi^i$s or $\eta^is$, we only need to find suitable matrix $A$. This allows one to systematically identify symmetry operations on the bosonic side \cite{po2021symmetric}.
In following sections we will discuss symmetry transformations in examples in detail.

\section{Majorana representation for square lattice}\label{sec_2}
In this section, we study the example of square lattice. After discussions of contents of operators,  we discuss how to express the states in terms of qubits. We also discuss some issues pertaining to the global properties of the transformation, e.g. issues of putting the system on a torus. We show local permutations of Majorana partons and rotational symmetry transformations explicitly.

Let us focus on the case of spin-$1/2$ fermions and a single orbital per site. On each site the Hilbert space is four-dimensional, spanned by the basis $\{\ket{0}, f^\dagger_\uparrow \ket{0}, f^\dagger_\downarrow \ket{0}, f^\dagger_\uparrow f^\dagger_\downarrow \ket{0}\}$. Generically the Hamiltonian of the system is built from quadratic forms of fermionic creation and annihilation operators. The products of quadratic forms compose interaction terms. So we mainly focus on quadratic terms. Let $\gamma^i, i=1,2,3,4$ be on-site Majorana operators such that:

\begin{equation}
\begin{split}
f_\uparrow &=\frac{1}{2}(\gamma^1-i\gamma^2);\quad f^\dagger_\uparrow=\frac{1}{2}(\gamma^1+i\gamma^2), \\
f_\downarrow &=\frac{1}{2}(\gamma^3-i\gamma^4);\quad f^\dagger_\downarrow=\frac{1}{2}(\gamma^3+i\gamma^4).\\
\end{split}
\end{equation}

\begin{figure}[htpb]
\begin{center}
	\begin{tikzpicture}
	 \foreach \x in {0,2}{
		\draw (\x, -0.5)--(\x, 0.5);
		\draw (-0.5, \x)--(0.5, \x);
		\draw (1.5, \x)--(2.5,\x);
		\draw (\x, 1.5)--(\x, 2.5);
		\node[above] at (-0.5,\x) {$3$}; \node[above] at (1.5,\x) {$3$};
		\node[below] at (0.5,\x) {$4$}; \node[below] at (2.5,\x) {$4$};
		\node[left] at (\x, -0.5) {$1$}; \node[left] at (\x, 1.5) {$1$};
		\node[right] at (\x, 0.6) {$2$}; \node[right] at (\x, 2.6) {$2$};
	}

		\draw[<->]  (-1.3,1.4)--(-1.3,1.0)--(-0.9,1.0);
		\node[right] at (-0.9,1.0){$\vec{\hat{x}}$};
		\node[above] at (-1.3,1.4){$\vec{\hat{y}}$};
		
		\foreach \x in {0,2}{
			\draw[thick] (-0.5, \x)--(0,\x+0.5);
			\draw[thick] (-0.5, \x)--(0,\x-0.5);
			\draw[thick] (0.5, \x)--(0,\x+0.5);
			\draw[thick] (0.5, \x)--(0,\x-0.5);
			\draw[thick] (1.5, \x)--(2,\x+0.5);
			\draw[thick] (1.5, \x)--(2,\x-0.5);
			\draw[thick] (2.5, \x)--(2,\x+0.5);
			\draw[thick] (2.5, \x)--(2,\x-0.5);
		}
	
	\foreach \x in {0,2}{
		\draw[->, dashed] (\x, 1.5)--(\x, 1.0);
		\draw[dashed] (\x, 1.0)--(\x,0.5);
		\draw[<-, dashed] (1.0, \x)--(1.5, \x);
		\draw[dashed] (0.5,\x)--(1.0,\x);
		\draw[->, dashed] (\x, -0.5)--(\x,-1.0);
		\draw[->, dashed] (-0.5,\x)--(-1.0,\x);
		\draw[->, dashed] (\x, 3)--(\x,2.7); \draw[dashed] (\x, 2.7)--(\x,2.5);
		\draw[->, dashed] (3,\x)--(2.7,\x); \draw[dashed] (2.7,\x)--(2.5,\x);
	}

	\end{tikzpicture}
\end{center}
	\caption{The auxiliary Majorana fermions of a square lattice. Each site has been attached a square, with every vertex $i$ representing $\chi^i_\vec{r}$. Each dashed line connects two lattice sites, and its arrow tells how to sign an intersite product $i\chi^i_\vec{r}\chi^j_\vec{r'}$.}
	\label{sq_lat}
\end{figure}
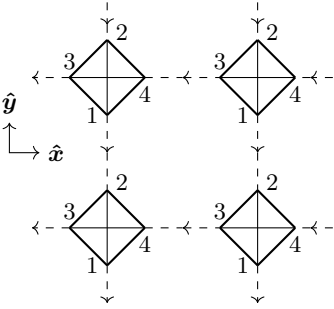

 For simplicity we choose arrow directions to be antiparallel to directions of basis vectors of the lattice, as shown in Fig. \ref{sq_lat}. Then by the definition in Eq. \ref{parton_cons} and the mapping in Eq. \ref{mapping} we obtain $\Theta^{ij}, \Lambda^{ij},\Phi^{ij},i,j\in\{1,2,3,4\}$. 
 
We can study some examples to understand the meaning of $\Theta_{\vec{r}}$ and $\Lambda_{\vec{r}}$ operators. First consider on-site particle number operator for spin up(similar for spin down),
\begin{equation}
n_\uparrow=f^\dagger_\uparrow f_\uparrow=\frac{1+i\gamma^2\gamma^1}{2}\rightarrow\frac{1+\Theta^{21}_{\vec{r}}}{2},
\end{equation}
If we replace $\Theta^{21}$ with its $\eta$ representation, then it turns out that $\eta^i$ has the similar physical meaning with $\gamma^i$. In other words, if we recombine $\eta$'s again into complex fermionic partons, say, 
\begin{equation}\begin{split}
&\eta^1=a+a^\dagger, \eta^2=i(a-a^\dagger), \\
&\eta^3=b+b^\dagger, \eta^4=i(b-b^\dagger),
\end{split}
\end{equation}
then $a^\dagger a$ and $b^\dagger b$ correspond to number of spin up and down particles respectively. We remark that although such equivalence is evident from the perspective of the on-site operators, $\eta^i_{\vec{r}}$ is nevertheless different from $\gamma^i_{\vec{r}}$ since $\gamma^i_{\vec{r}}\gamma^j_{\vec{r'}}\neq \eta^i_{\vec{r}}\eta^j_{\vec{r'}}$. This is where $\chi^i_{\vec{r}}$ plays a crucial role---to connect different sites. Actually, the parametrization of $\Theta$ and $\Lambda$ could be done in a different way. For this spinful fermions on square lattice, the Lie algebra formed by $\Theta^{ij}$ has an equivalence $\mathfrak{so}(4)\cong \mathfrak{su}(2)\oplus \mathfrak{su}(2)$; on Lie group level it is $Spin(4)\cong SU(2)\times SU(2)$. Using this, one could separate charge and spin degrees of freedom in the parton description and correspondence with the physical fermions will not be as simple \cite{po2021symmetric}.

We further introduce complex fermionic partons for $\chi$,
\begin{equation}\begin{split}
&\chi^1=d+d^\dagger, \chi^2=i(d-d^\dagger), \\
&\chi^3=g+g^\dagger, \chi^4=i(g-g^\dagger).
\end{split}
\end{equation}
Auxiliary fermion modes $d^\dagger$ and $g^\dagger$ enlarge the Hilbert space of a single site to be 16-dimensional. To obtain the correct physical Hilbert space we need some extra reductions. The discussion in \cite{po2021symmetric} requires the parton parity on-site to be frozen at $-1$, but such a constraint is not present here. Nevertheless, as the bosonized Hamiltonian is built using the $\Theta$, $\Lambda$, and $\Phi$ operators, which are all bilinears in the $\eta$ and $\chi$, the on-site parity operator for any site commutes with the bosonized Hamiltonian. In other words, we can project to a particular sector of on-site parton parity operator to obtain a bosonic model, i.e.
\begin{equation}
\Gamma_{\vec{r}}\equiv i^4 \eta^1_{\vec{r}}\eta^2_{\vec{r}}\eta^3_{\vec{r}}\eta^4_{\vec{r}}\chi^1_{\vec{r}}\chi^2_{\vec{r}}\chi^3_{\vec{r}}\chi^4_{\vec{r}}\ceq\rho,
\end{equation}
where $\rho\in \{+1,-1\}$ is a constant which we take to be $\vec r$-independent. ``$\ceq$'' means that the equality holds in the constrained Hilbert space.

\subsection{Qubit representation}

The on-site Hilbert space in the $\eta$-$\chi$ description is generated by acting creation operators $a^\dagger, b^\dagger, d^\dagger$ and $g^\dagger$ on the vaccum. We can introduce some qubits for each site, and identify an unoccupied state $\ket{0}$ with $\ket{\uparrow}$  and an occupied state $\ket{1}$ with $\ket{\downarrow}$, then the Hilbert space can be represented by 4 qubits. Let the basis states be defined as
\begin{equation}\label{state_sq}
\ket{n_a n_b n_d n_g}=(a^\dagger)^{n_a} (b^\dagger)^{n_b} (d^\dagger)^{n_d} (g^\dagger)^{n_g}\ket{0}.
\end{equation}
They correspond to 4-qubit basis naturally, for instance, $$\ket{1000}=\ket{\downarrow\uparrow\uparrow\uparrow}, \ket{0110}=\ket{\uparrow\downarrow\downarrow\uparrow}.$$
The actions of $\Theta_{\vec{r}}, \Lambda_{\vec{r}}$ and $\Phi_{\vec{r}}$ on the Hilbert space can be represented as strings of Pauli matrices and identity operator, $\{\mathbbm{1}, X, Y, Z\}$. For instance, 
\begin{equation}
\Theta^{12}=i^2(a+a^\dagger)(a-a^\dagger)=(1-2n_a)=Z^{(1)},
\end{equation}
where the superscript ``1'' refers to first qubit. The equality holds in the sense of acting on quantum states. The operators in qubit representation are as follows:
\begin{subequations}
\begin{equation}\label{theta_sq}
\begin{array}{ll}
\Theta^{12}=Z^{(1)}; &\Theta^{34}=Z^{(2)}; \\
\Theta^{13}=Y^{(1)}X^{(2)}; &\Theta^{24}=-X^{(1)}Y^{(2)}; \\
\Theta^{14}=-Y^{(1)}Y^{(2)}; &\Theta^{23}=X^{(1)}X^{(2)},\\
\end{array}
\end{equation}

\begin{equation}
\begin{array}{ll}
\Lambda^{11}=Y^{(1)}Z^{(2)}X^{(3)}; &\Lambda^{22}=-X^{(1)}Z^{(2)}Y^{(3)}; \\
\Lambda^{33}=Y^{(2)}Z^{(3)}X^{(4)}; &\Lambda^{44}=-X^{(2)}Z^{(3)}Y^{(4)}; \\ 
\end{array}
\end{equation}

\begin{equation}\label{sq_phi}
\begin{array}{ll}
\Phi^{12}=-Z^{(3)}; &\Phi^{34}=-Z^{(4)}; \\
\Phi^{13}=-Y^{(3)}X^{(4)}; &\Phi^{24}=X^{(3)}Y^{(4)}; \\
\Phi^{14}=Y^{(3)}Y^{(4)}; &\Phi^{23}=-X^{(3)}X^{(4)},\\
\end{array}
\end{equation}
\end{subequations}
We can, however, impose the projection  $\Gamma_{\vec{r}}=Z^{(1)}_\vec{r}Z^{(2)}_\vec{r}Z^{(3)}_\vec{r}Z^{(4)}_\vec{r}\ceq\rho$ to reduce the degrees of freedom by half. Equivalently, the last qubit $\sups{Z}{4}$ is in fact fully determined by the other three qubits.
We can therefore obtain a more efficient description using only the first three qubit, with the state in the last qubit understood to be constrained by that of the first three. This way, we can simply replace the Pauli operators on the $4^{\rm th}$ qubit by the coefficient they generate, i.e.,

\begin{equation}\label{hidden}
\begin{split}
&X^{(4)}\Rightarrow \mathbbm{1}; \\
&Z^{(4)}\Rightarrow \rho Z^{(1)} Z^{(2)} Z^{(3)}; \\
&Y^{(4)}\Rightarrow i\rho Z^{(1)} Z^{(2)} Z^{(3)}, \\
\end{split}
\end{equation}
The last line can also be obtained from $\sups{Y}{4}=i\sups{X}{4}\sups{Z}{4}$.
By these replacements the last qubit is ``hidden'' while the possible coefficients from actions on the last qubit are absorbed into operators acting on other qubits. Then $\Lambda^{ij}$ is a $4\times 4$ operator-valued matrix 

\begin{widetext}
\begin{equation}
\Lambda^{ij}=
\left(
\begin{array}{cccc}
Y^{(1)}Z^{(2)}X^{(3)} & -Y^{(1)}Z^{(2)}Y^{(3)} & Y^{(1)}Z^{(2)}Z^{(3)} &\rho X^{(1)}\\
X^{(1)}Z^{(2)}X^{(3)} & -\sups{X}{1}\sups{Z}{2}\sups{Y}{3} & \sups{X}{1}\sups{Z}{2}\sups{Z}{3} & -\rho \sups{Y}{1}\\
\sups{Y}{2}\sups{X}{3} & -\sups{Y}{2}\sups{Y}{3} & \sups{Y}{2}\sups{Z}{3} & \rho \sups{Z}{1}\sups{X}{2} \\
\sups{X}{2}\sups{X}{3} & -\sups{X}{2}\sups{Y}{3} & \sups{X}{2}\sups{Z}{3} & -\rho\sups{Z}{1}\sups{Y}{2} \\
\end{array}
\right)_{ij} .
\end{equation}
\end{widetext}

Comparing to fermionic Hilbert space, we have an extra qubit as an auxiliary degree of freedom. This unphysical degree of freedom is consumed when we consider the mapping of an identity from the fermionic side:

\begin{equation}\label{con_sq}
\begin{split}
1&=(i\gamma_\vec{r}^2 \gamma_\vec{r}^4) (i\gamma_\vec{r}^4\gamma^3_{\vec{r}+\vec{x}})
(i \gamma^3_{\vec{r}+\vec{x}} \gamma_{\vec{r}+\vec{x}}^2) (i\gamma_{\vec{r}+\vec{x}}^2 \gamma_{\vec{r}+\vec{x}+\vec{y}}^1) \\ 
&\times(i\gamma_{\vec{r}+\vec{x}+\vec{y}}^1 \gamma_{\vec{r}+\vec{x}+\vec{y}}^3) (i\gamma_{\vec{r}+\vec{x}+\vec{y}}^3 \gamma_{\vec{r}+\vec{y}}^4) 
(i \gamma_{\vec{r}+\vec{y}}^4\gamma_{\vec{r}+\vec{y}}^1)
(i \gamma_{\vec{r}+\vec{y}}^1\gamma^2_\vec{r})\\
&\rightarrow 1=\Theta^{24}_\vec{r} \Lambda^{44}_\vec{r}\Lambda^{33}_{\vec{r}+\vec{x}} \Theta^{32}_{\vec{r}+\vec{x}}\Lambda^{22}_{\vec{r}+\vec{x}}\Lambda_{\vec{r}+\vec{x}+\vec{y}}^{11} \Theta^{13}_{\vec{r}+\vec{x}+\vec{y}} \\ &\qquad\qquad\times\Lambda^{33}_{\vec{r}+\vec{x}+\vec{y}}\Lambda^{44}_{\vec{r}+\vec{y}} \Theta^{41}_{\vec{r}+\vec{y}}\Lambda^{11}_{\vec{r}+\vec{y}}\Lambda^{22}_\vec{r} \\
&\Rightarrow \hat{C}_{\vec{r}}\equiv \Phi^{24}_\vec{r}\Phi^{32}_{\vec{r}+\vec{x}}\Phi^{13}_{\vec{r}+\vec{x}+\vec{y}} \Phi^{41}_{\vec{r}+\vec{y}} \ceq -1.
\end{split}
\end{equation}

As discussed above, Majorana operators $\chi^i$ are auxiliary and ``unphysical'', so $\Phi^{ij}$s as bilinears of auxiliary operators somehow play the role of "gauge operators". Eq. \ref{con_sq} is a constraint such that auxiliary degrees of freedom are restricted so as to be consistent with the fermion identity on the first line. For each plaquette with left-bottom corner at $\vec{r}$ there is a constraint equation, so effectively there is one constraint for each site thus the on-site Hilbert space of the system is reduced to 4-dimensional. These constraints can be implemented in Hamiltonian as a summation $K\sum_{\vec{r}}\hat{C}_{\vec{r}}$ with coupling $K$ sufficient large so that the $\hat{C}_{\vec{r}}$ are enforced to be $-1$.

In qubit representation, by exploiting expressions of $\Phi^{ij}$ the constraint can be shown as:
\begin{equation}
\sups{Y}{3}_{\vec{r}}\sups{X}{3}_{\vec{r}+\vec{x}}\sups{Y}{3}_{\vec{r}+\vec{x}+\vec{y}} \sups{X}{3}_{\vec{r}+\vec{y}}=(\sups{Z}{1}\sups{Z}{2})_{\vec{r}} (\sups{Z}{1}\sups{Z}{2})_{\vec{r}+\vec{y}}
\end{equation}

The left hand side of the constraint is the Hamiltonian in Wen's plaquette model \cite{PhysRevLett.90.016803}, which is known to be equivalent to toric code model \cite{KITAEV20032} and describes the $\mathbb{Z}_2$ topological order. In the work of Chen etc. such constraints are interpreted as flux attachments of gauge fields \cite{CHEN2018234, PhysRevB.100.245127,  PhysRevResearch.2.033527}. In our construction, the left hand side is not fixed but depends on values of physical degrees of freedom.

\subsection{Wilson loops and fermionic odd operators \label{sec:graph_count}}

If we put the lattice system on a torus manifold, i.e. with periodic boundary conditions, then there will be extra global constraints from Wilson loops. To see this let lattice size be $L_x\times L_y$ and $r_{x,y}\in\mathbb{Z}/(L_{x,y}\mathbb{Z})$. From fermionic side, there is an identity for products of Majorana operators along a $y$-loop, 
\begin{equation}
(i\gamma_\vec{r}^1\gamma_\vec{r}^2)(i\gamma_\vec{r}^2\gamma_{\vec{r}+\vec{y}}^1)\cdots(i\gamma_{\vec{r}-\vec{y}}^2\gamma_\vec{r}^1)=(-1)^{L_y}.
\end{equation}
 Mapped to bosonic side, it becomes a constraint:
\begin{equation}
W_y(\vec{r})\equiv \Phi^{12}_\vec{r}\Phi_{\vec{r}+\vec{y}}^{12}\cdots\Phi^{12}_{\vec{r}-\vec{y}}=-1.
\end{equation}
Notice that this constraint is actually independent of the choice of base point $\vec{r}$. Topologically the $y$-loop is a homology class, for which various deformations can be achieved by plaquette constraints(Eq. \ref{con_sq}) in last section. Along $x$-direction there is also a similar constraint
\begin{equation}
W_x(\vec{r})\equiv \Phi^{34}_\vec{r}\Phi_{\vec{r}+\vec{x}}^{34}\cdots\Phi^{34}_{\vec{r}-\vec{x}}=-1.
\end{equation}

From two Wilson loop constraints we obtain a global constraint of fermion parity:
\begin{equation}\label{sq_fp}
\begin{split}
\hat{P} &\equiv\rho^{L_xL_y}\prod_{n=0}^{L_x-1}W_{y}(\vec{r}+n\vec{x})\prod_{n=0}^{L_y-1}W_{x}(\vec{r}+n\vec{y}) \\
&=\rho^{L_xL_y}\prod_{m=0}^{L_x-1}\prod_{n=0}^{Ly-1}\Phi^{12}_{m\vec{x}+n\vec{y}}\Phi^{34}_{m\vec{x}+n\vec{y}}\\
&=\prod_{m=0}^{L_x-1}\prod_{n=0}^{Ly-1}\Theta^{12}_{m\vec{x}+n\vec{y}}\Theta^{34}_{m\vec{x}+n\vec{y}}\\
&\ceq (-1)^{L_x+L_y}\rho^{L_xL_y},
\end{split}
\end{equation}
where we have used the parton parity on each site, $\Gamma_{\vec{r}}=\Theta_{\vec{r}}^{12}\Theta_{\vec{r}}^{34}\Phi^{12}_\vec{r}\Phi^{34}_\vec{r}=\rho$. On fermionic side, $\Theta^{12}_\vec{r}\Theta^{34}_\vec{r}$ is equal to on-site fermion parity $(-1)^{n_\vec{r}}$, so the product over whole lattice gives global fermionic parity. Combining Wilson loop constraints and on-site parton parity fixing, we can describe half of the physical Hilbert space where fermion parity of states are fixed. For instance, to make the vacuum state parity even, $L_x$ and $L_y$ should be both odd or both even. In former case $\rho$ has to be $+1$ while in latter case $\rho$ can be either $+1$ or $-1$. 

To describe the whole Hilbert space, we can consider bosonization of fermion-odd operator, like $\gamma^{i}_\vec{r}$. The basic idea is, since we have two Wilson loops, we may modify the fermion-boson operator mapping such that one of the Wilson loop constraints is violated. This can be implemented by introducing a ``defect'' in a certain site, making an arrow of one its link reversed. Then by checking the commutation relations one can regard one $\Lambda^{ii}_\vec{r}$ as bosonization of $\gamma^i_{\vec{r}}$ after such manipulations. This is similar to the case where we choose a starting point in 1D Jordan Wigner transformation so that fermionic-odd opeartors are mapped to Pauli strings. For square lattice this has been discussed in detail in \cite{po2021symmetric}. We do not show  details temporarily, but will discuss this in Sec.\ref{fermion-odd_cub} for cubic lattices.

\subsection{Local Permutations and Symmetries}

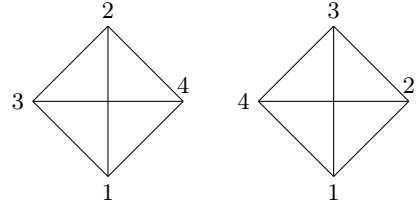
\begin{figure}[htpb]
	\begin{tikzpicture}

	\foreach \x in {0,3}{
	\draw (\x,-1)--(\x,1);
	\draw (\x-1,0)--(\x+1,0);
	\draw (\x-1,0)--(\x,1); \draw (\x-1,0)--(\x,-1);
	\draw (\x+1,0)--(\x,1); \draw (\x+1,0)--(\x,-1);
	}

	\node[left] at(-1,0){3};
	\node[above] at(1,0){4};
	\node[below] at(0,-1){1};
	\node[above] at(0,1) {2};
	
	\node[left] at(2,0){4};
	\node[above] at(4,0){2};
	\node[below] at(3,-1){1};
	\node[above] at(3,1) {3};
	\end{tikzpicture}
	\caption{Two choices of labeling. The right choice can be obtained by doing local unitary transformation.
	\label{labeling_choices}}
\end{figure}

In the setup above, we have labelled the four auxiliary Majorana $\chi^i$s in an antipodal way. One may ask if there is any preference in choosing a certain labeling order, in other words, what is the connection and difference between two artificial choices(see Fig. \ref{labeling_choices} for example). It turns out that different choices of labelling are physically equivalent by local unitary transformations. To see this, we look for a SO(4) rotation transformation of $\chi^i$ from one labeling order to another as discussed in Sec. \ref{general_procedures}. For instance, the transformation in Fig. \ref{labeling_choices} is:
\begin{equation}
	(\Lambda^{i1} \ \Lambda^{j2} \ \Lambda^{k3} \ \Lambda^{l4})\Rightarrow 
	(\Lambda^{i1} \ \Lambda^{j2} \ \Lambda^{k3} \ \Lambda^{l4})\left(
	\begin{array}{cccc}
	1 & 0 & 0 & 0 \\
	0 & 0 & 0 & 1 \\
	0 & 1 & 0 & 0 \\
	0 & 0 & 1 & 0 \\
	\end{array}
	\right).
\end{equation}
The transformation matrix is a permutation operation with determinant 1(if determinant is -1 then one needs to add a minus sign before one non-vanishing entry to preserve orientation), and thus is an element in group SO(4). Generically we can find an anti-symmetric real matrix $A$ such that the transformation matrix is equal to $e^{-2A}$, then an unitary matrix $V(A)=e^{-i\sum_{ij}\phi^{ij}A_{ij}}$ will permute the second index of $\Lambda^{ij}$. In the example above, the solution is,
\begin{subequations}
\begin{equation}
	A=\frac{-1}{2}\left(
	\begin{array}{cccc}
	0 & 0 & 0 & 0 \\
	0 & 0 & -\frac{2 \pi }{3 \sqrt{3}} & \frac{2 \pi }{3 \sqrt{3}} \\
	0 & \frac{2 \pi }{3 \sqrt{3}} & 0 & -\frac{2 \pi }{3 \sqrt{3}} \\
	0 & -\frac{2 \pi }{3 \sqrt{3}} & \frac{2 \pi }{3 \sqrt{3}} & 0 \\
	\end{array}
	\right)
\end{equation}

\begin{equation}
V(A)=e^{-i\frac{2\pi}{3\sqrt{3}}\left( -\phi^{23}+\phi^{24}-\phi^{34}\right) }
\end{equation}
\end{subequations}

 The example above does not touch physical symmetries. When we consider internal unitary symmetries, like charge conjugation, particle hole symmetry, we can use $U(A)$ and $V(A)$ with different $A$ matrices. For spatial symmetry transformation, besides rotating the lattice sites(we call it ``bare rotation''), we should also rotate both labels of $\chi$s and intersite arrows correspondingly. For example, the $C_4$ rotation transformation can be represented by combination of bare rotation $C_4^{b}$ and internal unitary $V_{C_4}$: 
\begin{subequations}\label{sq_rotation}
\begin{equation}
C_4=V_{C_4}C_4^{b};
\end{equation}
\begin{equation}
\begin{split}
&C^{b}_4\Lambda^{44}_\vec{r} \Lambda^{33}_{\vec{r}+\vec{x}}(\hat{C}_4^{b})^{-1}=\Lambda^{44}_\vec{r}\Lambda^{33}_{\hat{C}_4\vec{r}+y}, \\
&V_{C_4}=e^{-i\frac{\pi}{4}\sum_\vec{r}\left( \phi^{12}_\vec{r} +\phi^{34}_\vec{r}-\sqrt{2} (\phi^{13}_\vec{r}+\phi^{24}_\vec{r}+\phi^{14}_\vec{r}-\phi^{23}_\vec{r}) \right)},
\end{split}
\end{equation}
\end{subequations}
such that 
\begin{equation}
	(\Lambda^{i1} \ \Lambda^{j2} \ \Lambda^{k3} \ \Lambda^{l4})\Rightarrow 
(-\Lambda^{i4} \ \Lambda^{j3} \ \Lambda^{k1} \ \Lambda^{l2}),
\end{equation}
where we have added a ``$-$'' sign so as to preserve the directions of arrows. For suqare lattice more details of these transformations can be found in \cite{po2021symmetric}. In this sense, our bosonization strategy makes symmetries of lattice system manifest. 

To conclude this section, we compare the present $\eta\chi$ construction with the approach in Ref. \onlinecite{po2021symmetric}. In Ref. \onlinecite{po2021symmetric}, the parton construction is designed to implement a sense of  spin-charge separation, so partons therein have different physical meaning. This can be regarded as choosing different basis for the Clifford algebra generated by $\eta^i$ and $\chi^i$. In Appendix \ref{irreps}, we show that qubit representation here can be turned into qubit representation Eq. (112) in Ref. \onlinecite{po2021symmetric} by a local unitary transformation. We also show that after imposing the parity constraint on the partons, the on-site Hilbert space for both approaches furnishes a spinor representation of SO(8), and states with the same weight on the two sides correspond to the same fermionic state.

\section{Triangular lattice}\label{sec_3}

Our second example is bosonization of triangular lattice system. Each vertex has coordination number six.  We will consider spinful fermion, although it can be generalized to arbitrary number of flavors.

For a triangular lattice, on each site there are six links, so similar to square lattice case we start from $\chi^i, i=1,2,...,6$ and $\eta^i,i=1,2,3,4$.  The on-site Hilbert space is now $2^5=32$ dimensional. Based on square lattice case, we extend complex fermion representation to $\chi^5,\chi^6$:
\begin{equation}\begin{split}
&\chi^5=h+h^\dagger, \chi^6=i(h-h^\dagger).
\end{split}
\end{equation}
The the basis states of on-site Hilbert space are in the form:
\begin{equation}\label{state_tri}
\ket{n_a n_b n_c n_d n_g n_h}=(a^\dagger)^{n_a} (b^\dagger)^{n_b} (d^\dagger)^{n_d} (g^\dagger)^{n_g}(h^\dagger)^{n_h}\ket{0}.
\end{equation}
On each site we also define a parton parity operator and make projection to its eigenspace with eigenvalue $\rho$,
\begin{equation}
\begin{split}
\Gamma_{\vec{r}} &\equiv -i^5 \eta^1_{\vec{r}}\eta^2_{\vec{r}}\eta^3_{\vec{r}}\eta^4_{\vec{r}}\chi^1_{\vec{r}}\chi^2_{\vec{r}}\chi^3_{\vec{r}}\chi^4_{\vec{r}}\chi^5_{\vec{r}}\chi^6_{\vec{r}}\ceq \rho, \\
&\Leftrightarrow \Theta_{\vec{r}}^{12}\Theta_{\vec{r}}^{34}\Phi_{\vec{r}}^{12} \Phi_{\vec{r}}^{34} \Phi_{\vec{r}}^{56}=\rho
 \end{split}
\end{equation}

\begin{figure}[htpb]
	\begin{tabular}{m{4cm} m{4cm}}
		\begin{tikzpicture}[global scale = 0.75]
		\tikzmath{ \r = 1; };
		\path[coordinate] (0,0)  coordinate(A)
		++( 120:1cm) coordinate(B)
		++(60:1cm) coordinate(C)
		++(0:1cm) coordinate(D)
		++(-60:1cm) coordinate(E)
		++(240:1cm) coordinate(F)
		++(120:1cm) coordinate(G)
		;
		\draw (A) -- (B) -- (C) --(D) -- (E) -- (F)-- cycle;
		\node[left] at (A) {4};
		\node[left] at (B) {6};
		\node[left] at (C) {1};
		\node[below left] at (D) {3};
		\node[left] at (E) {5};
		\node[above left] at (F) {2};
		\node at (60:1cm) {$\vec{r}$};

		\draw[-<, dashed] (C)-- ++(120:0.5) coordinate(X);
		\draw[dashed] (X)-- ++(120: 0.5);
	
		\path[coordinate] (3.0,0)  coordinate(A)
		++( 120:1cm) coordinate(B)
		++(60:1cm) coordinate(C)
		++(0:1cm) coordinate(D)
		++(-60:1cm) coordinate(E)
		++(240:1cm) coordinate(F)
		++(120:1cm) coordinate(G)
		;
		\draw (A) -- (B) -- (C) --(D) -- (E) -- (F)-- cycle;
		\draw[dashed] (1.5, 0.866) -- (2.25, 0.866);
		\draw[<-, dashed] (2.0, 0.866) -- (2.5, 0.866);
		\node at (G) {$\vec{r}+\vec{a}$};
		
		\path[coordinate] (1.5, 2.598)  coordinate(A)
		++( 120:1cm) coordinate(B)
		++(60:1cm) coordinate(C)
		++(0:1cm) coordinate(D)
		++(-60:1cm) coordinate(E)
		++(240:1cm) coordinate(F)
		++(120:1cm) coordinate(G)
		;
		\draw (A) -- (B) -- (C) --(D) -- (E) -- (F)-- cycle;
		\node at (G) {$\vec{r}+\vec{b}$};

		\draw[-<, dashed] (C)-- ++(120:0.5) coordinate(X);
		\draw[dashed] (X)-- ++(120: 0.5);
		\draw[->, dashed] (B)-- ++(180:0.5) coordinate(X);
		\draw[dashed] (X)-- ++(180: 0.5);
		\draw[-<, dashed] (D)-- ++(60:0.5) coordinate(X);
		\draw[dashed] (X)-- ++(60: 0.5);
		\draw[-<, dashed] (E)-- ++(0:0.5) coordinate(X);
		\draw[dashed] (X)-- ++(0: 0.5);
		
		\draw[dashed] (1.0, 1.732) -- (1.25, 2.165);
		\draw[<-, dashed] (1.25, 2.165) -- (1.5, 2.598);
		
		\draw[dashed] (3.0, 1.732) -- (2.75, 2.165);
		\draw[->, dashed] (2.5, 2.598)--(2.75, 2.165);
		\end{tikzpicture}
		&
		\begin{tikzpicture}
		\tikzmath{ \rows = 3; };

		\foreach \row in {0, 1, ...,\rows} {
			\draw ($\row*(0.5, {0.5*sqrt(3)})$) -- ($(\rows,0)+\row*(-0.5, {0.5*sqrt(3)})$);
			\draw ($\row*(1, 0)$) -- ($(\rows/2,{\rows/2*sqrt(3)})+\row*(0.5,{-0.5*sqrt(3)})$);
			\draw ($\row*(1, 0)$) -- ($(0,0)+\row*(0.5,{0.5*sqrt(3)})$);
		};
		
		\foreach \row in {0, 1, ...,\rows} {
			\draw ($\row*(0.5, -{0.5*sqrt(3)})+(\rows/2,{\rows/2*sqrt(3)})$) -- ($(\rows,0)-\row*(0.5,{0.5*sqrt(3)})+(\rows/2,{\rows/2*sqrt(3)})$);
			\draw ($\row*(1, 0)+(\rows/2,{\rows/2*sqrt(3)})$) -- ($(\rows,0)+\row*(0.5,{0.5*sqrt(3)})$);
			\draw ($\row*(1, 0)+(\rows/2,{\rows/2*sqrt(3)})$) -- ($(\rows/2,{\rows/2*sqrt(3)})+\row*(0.5,{-0.5*sqrt(3)})$);
		};
		\draw[->, thick, red] (0,0)--($(0.5, {0.5*sqrt(3)})$) ;
		\draw[->, thick, red] (0,0)--(1,0) ;
		\node[left] at ($(0.5, {0.5*sqrt(3)})$) {$\vec{b}$};
		\node[below] at (1,0){$\vec{a}$};
		\end{tikzpicture}
	\end{tabular}
	\caption{The right figure is a triangular lattice with two basis vectors in red. The left figure shows that six auxiliary Majorana fermions are attached to each site, with labels 1 to 6.}
\end{figure}
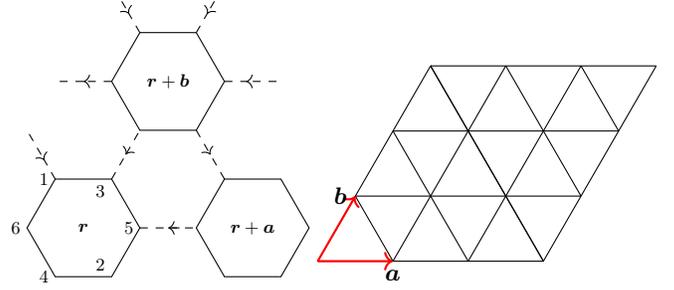

\subsection{Qubit representation}

As shown in Eq. \ref{state_tri}, we can use five qubits to represent the states similar to Eq. \ref{state_sq}. For $i,j\in\{1,2,3,4\}$ the expressions of $\Theta^{ij}$ are the same as Eq. \ref{theta_sq}. For other operators, one should notice that anticommutation relations bring extra Paul $Z$ operators. 

Parity operator is $\Gamma_\vec{r}=\sups{Z}{1}\sups{Z}{2}\sups{Z}{3}\sups{Z}{4} \sups{Z}{5} \sups{Z}{6}$. We fix $\Gamma_\vec{r}=\rho\in\{+1,-1\}$ so that the last qubit can be hidden as in Eq. \ref{hidden}. Operations acting on the last qubit can be replaced by:
\begin{align}\label{replace_6}
\begin{split}
&X^{(5)}\Rightarrow \mathbbm{1} \\
&Z^{(5)}\Rightarrow \rho Z^{(1)} Z^{(2)} Z^{(3)} \sups{Z}{4}\\
&Y^{(5)}\Rightarrow i\rho Z^{(1)} Z^{(2)} Z^{(3)}  \sups{Z}{4}\\
\end{split}
\end{align}
Then $\Lambda^{ij}$ matrix is
\begin{widetext}
\begin{equation}
\Lambda^{ij}=
\left(
\begin{array}{cccccc}
Y^{(1)}Z^{(2)}X^{(3)} & -Y^{(1)}Z^{(2)}Y^{(3)} & Y^{(1)}Z^{(2)}Z^{(3)}\sups{X}{4} &-Y^{(1)}Z^{(2)}\sups{Z}{3}Y^{(4)} &Y^{(1)}Z^{(2)}Z^{(3)}\sups{Z}{4} &\rho X^{(1)}\\
X^{(1)}Z^{(2)}X^{(3)} & -\sups{X}{1}\sups{Z}{2}\sups{Y}{3} & \sups{X}{1}\sups{Z}{2}\sups{Z}{3}\sups{X}{4}
&-\sups{X}{1}\sups{Z}{2}\sups{Y}{3}\sups{Y}{4} &X^{(1)}Z^{(2)}\sups{Z}{3}\sups{Z}{4}
&-\rho \sups{Y}{1}\\
\sups{Y}{2}\sups{X}{3} & -\sups{Y}{2}\sups{Y}{3} & \sups{Y}{2}\sups{Z}{3}\sups{X}{4} 
&-\sups{Y}{2}\sups{Z}{3}\sups{Y}{4}
&\sups{Y}{2}\sups{Z}{3}\sups{Z}{4}
&\sups{Z}{1}\sups{X}{2}\\
\sups{X}{2}\sups{X}{3} & -\sups{X}{2}\sups{Y}{3} & \sups{X}{2}\sups{Z}{3}\sups{X}{4} 
& -\sups{X}{2}\sups{Z}{3}\sups{Y}{4}
&\sups{X}{2}\sups{Z}{3}\sups{Z}{4}
& -\rho \sups{Z}{1}\sups{Y}{2}\\
\end{array}
\right)_{ij} .
\end{equation}
\end{widetext}

We also list extra $\Phi$ operators besides Eq. \ref{sq_phi} for later use.

\begin{equation}
\begin{array}{ll}
\Phi^{56} =-\sups{Z}{5}; & \quad \\
\Phi^{15} = \sups{Y}{3}\sups{Z}{4}\sups{X}{5} &
\Phi^{16} =\sups{Y}{3}\sups{Z}{4}\sups{Y}{5};\\ 
\Phi^{25} =-\sups{X}{3}\sups{Z}{4}\sups{X}{5}; & \Phi^{26}=\sups{X}{3}\sups{Z}{4}\sups{Y}{5};\\
\Phi^{35}=-\sups{Y}{4}\sups{X}{5} &
\Phi^{36}=\sups{Y}{4}\sups{Y}{5}; \\
\Phi^{45}=-\sups{X}{4}\sups{X}{5}; &
\Phi^{46}=\sups{X}{4}\sups{Y}{5}.\\
\end{array}
\end{equation}

We turn to plaquette constraints. For triangular lattice there are two types of inequivalent plaquettes, which we denote as I and II. For instance, type I plaquette gives constraint:
\begin{equation}\begin{split}
-1 &=(i\gamma^{i_1}_\vec{r}\gamma^{i_2}_\vec{r}) (i\gamma^{i_2}_\vec{r}\gamma^{i_3}_{\vec{r}+\vec{a}}) (i\gamma^{i_3}_{\vec{r}+\vec{a}}\gamma^{i_4}_{\vec{r}+\vec{a}}) (i\gamma^{i_4}_{\vec{r}+\vec{a}}\gamma^{i_5}_{\vec{r}+\vec{b}}) \\
&\quad \times (i\gamma^{i_5}_{\vec{r}+\vec{b}}\gamma^{i_6}_{\vec{r}+\vec{b}}) (i\gamma^{i_6}_{\vec{r}+\vec{b}}\gamma^{i_1}_\vec{r}) \\
\rightarrow &\quad \hat{C}^{\text{I}}_\vec{r}\equiv\Phi^{35}_\vec{r} \Phi^{61}_{\vec{r}+\vec{a}} \Phi^{24}_{\vec{r}+\vec{b}} \ceq 1,
\end{split}
\end{equation}
where we choose $i_1\neq i_2, i_3\neq i_4, i_5\neq i_6$, and the superscript ``I'' for $\hat{C}^{\text{I}}_\vec{r}$ standing for type I plaquette. Similarly one can obtain constraint for type II plaquette:
\begin{equation}
\hat{C}^{\text{II}}_\vec{r}\equiv\Phi^{52}_\vec{r} \Phi^{13}_{\vec{r}+\vec{a}-\vec{b}} \Phi^{46}_{\vec{r}+\vec{a}} \ceq -1,
\end{equation}

Use qubit representation of $\Phi^{ij}$ and replacements Eq. \ref{replace_6}, we obtain constraints in terms of Pauli matrices:
\begin{equation}
\begin{split}
& \text{type I}: \sups{Y}{4}_\vec{r}\sups{X}{3}_{\vec{r}+a}\sups{X}{3}_{\vec{r}+\vec{b}} \sups{Y}{4}_{\vec{r}+\vec{b}}=-\rho \left(\sups{Z}{1}\sups{Z}{2} \right)_{\vec{r}+\vec{a}}\\
& \text{type II}: \sups{X}{3}_\vec{r}\sups{Z}{4}_\vec{r} \sups{Z}{3}_{\vec{r}+\vec{a}}\sups{Y}{4}_{\vec{r}+a}
\sups{Y}{3}_{\vec{r}+\vec{a}-\vec{b}} \sups{X}{4}_{\vec{r}+\vec{a}-\vec{b}}\\
&\qquad=\rho \left(\sups{Z}{1}\sups{Z}{2}\right)_{\vec{r}+\vec{a}}\\
\end{split}
\end{equation}

We can also combine these two constraints, up-down and left-right, as noted in Fig. \ref{combined_cons}. The second equation in the figure shows again similarity to Wen's plaquette model \cite{PhysRevLett.90.016803}, if we consider all states in eigenstates of Pauli $Z$.

A simple counting of degrees of freedom is as follows. Each plaquette constraint is shared by three sites, so it contributes $\frac{1}{3}$ constraints for each site. Then each site effectively has $6\times\frac{1}{3}=2$ plaquette constraints. Combining with the local parton parity projection, we get back to physical Hilbert space. 

\begin{figure}[htpb]
\begin{tikzpicture}
\draw [ultra thick] (0,0) to (1/2,{sqrt(3)/2} ) to (1,0) to (1/2, {-sqrt(3)/2} ) to (0, 0);
\draw [ultra thick] (0,0) to (1,0);
\node [left] at (0,0) {$\sups{X}{3}\sups{X}{4}$};
\node [right] at (1,0) {$\sups{Y}{3}\sups{Y}{4}$};
\node [above] at (1/2,{sqrt(3)/2} ) {$\sups{X}{3}\sups{Y}{4}$};
\node [below] at (1/2,{-sqrt(3)/2} ) {$\sups{Y}{3}\sups{X}{4}$};
\node [right] at (2.5,0) {$= (-\mathbbm{1} )\times$ };
\draw [ultra thick] (0+4,0) to (1/2+4,{sqrt(3)/2} ) to (1+4,0) to (1/2+4, {-sqrt(3)/2} ) to (0+4, 0);
\draw [ultra thick] (0+4,0) to (1+4,0);

\end{tikzpicture}

\begin{tikzpicture}
\draw [ultra thick] (0,0) to (1/2,{sqrt(3)/2} ) to (1,0) to (0, 0);
\draw [ultra thick] (0,0) to (-1/2,{sqrt(3)/2} ) to (1/2,{sqrt(3)/2} );
\node [left] at (0,0) {$\sups{Y}{3}\sups{Z}{4}$};
\node [right] at (1,0) {$\sups{X}{3}$};
\node [above right] at (1/2,{sqrt(3)/2} ) {$\sups{Y}{3}$};
\node [above] at (-1/2,{sqrt(3)/2} ) {$\sups{X}{3}\sups{Z}{4}$};
\node [right] at (2,0) {$=$};
\draw [ultra thick] (0+3,0) to (1/2+3,{sqrt(3)/2} ) to (1+3,0) to (0+3, 0);
\draw [ultra thick] (0+3,0) to (-1/2+3,{sqrt(3)/2} ) to (1/2+3,{sqrt(3)/2} );
\node [right] at (1+3,0) {$\rho\sups{Z}{1}\sups{Z}{2}$};
\node [above right] at (1/2+3,{sqrt(3)/2} ) {$\rho\sups{Z}{1}\sups{Z}{2}$};

\end{tikzpicture}
\caption{The combined constraints}
\label{combined_cons}
\end{figure}

\subsection{Wilson loops}
For triangular lattice with periodic boundary conditions, we can also understand the system as if they are on a torus manifold. Reversely it may be regarded as a triangulation of  a torus. Suppose the system size is $L_a\times L_b$, we have extra two Wilson loop constraints as follows. Along $\vec{a}$ direction, we have

\begin{equation}
W_a(\vec{r})\equiv \Phi^{65}_{\vec{r}}\Phi^{65}_{\vec{r}+\vec{a}} \cdots \Phi^{65}_{
\vec{r}-\vec{a}}\ceq -1.
\end{equation}
Similarly along $\vec{b}$ direction, 
\begin{equation}
W_b(\vec{r})\equiv \Phi^{43}_{\vec{r}}\Phi^{43}_{\vec{r}+\vec{b}} \cdots \Phi^{43}_{
	\vec{r}-\vec{b}}\ceq -1.
\end{equation}
There is a remaining type of loops along $(\vec{a}-\vec{b})$ direction, which can be expressed as products of Wilson loops along $\vec{a}$ and $\vec{b}$ directions and some plaquettes. 

One may wonder if the system is overconstrained. In fact, in periodic boundary case the product of all plaquette constraints is equal to identity, meaning one of them is dependent on others. So finally we receive a half Hilbert space with fixed global fermion parity, just like discussions below Eq. \ref{sq_fp}.

\subsection{Symmetries}
Besides translation symmetry along $\vec{a}, \vec{b}$ directions, triangular lattice also has dihedral group symmetry $D_6$ generated by a $60^\circ$ rotation $C_6$, and two reflections. $C_6$ rotation symmetry can be represented by bare rotation $C_6^b$ and local unitary transformation.  Bare $C_6$ rotation
\begin{equation}
C_6^b: \Lambda^{ij}_\vec{r}\Rightarrow \Lambda^{ij}_{C_6\vec{r}},
\end{equation}
and internal unitary transformation permutes $\chi$ indices 
\begin{equation}
	(\Lambda^{i1} \ \Lambda^{j2} \ \Lambda^{k3} \ \Lambda^{l4} \ \Lambda^{m5} \ \Lambda^{n6})\Rightarrow 
(\Lambda^{i6} \ -\Lambda^{j5} \ \Lambda^{k1} \ \Lambda^{l2} \ \Lambda^{m3} \ \Lambda^{n4})
\end{equation}
This can be realised by $V_{C_6}= e^{-i\pi/6\sum_{\vec{r}} \phi^{ij}\tilde{A}_{ij}}$, where 
\begin{equation}
\tilde{A}=\frac{1}{2}\left(
\begin{array}{cccccc}
0 & 1 & -2 & -\frac{2}{\sqrt{3}} & \frac{2}{\sqrt{3}} & 2 \\
-1 & 0 & \frac{2}{\sqrt{3}} & -2 & -2 & \frac{2}{\sqrt{3}} \\
2 & -\frac{2}{\sqrt{3}} & 0 & 1 & -2 & -\frac{2}{\sqrt{3}} \\
\frac{2}{\sqrt{3}} & 2 & -1 & 0 & \frac{2}{\sqrt{3}} & -2 \\
-\frac{2}{\sqrt{3}} & 2 & 2 & -\frac{2}{\sqrt{3}} & 0 & 1 \\
-2 & -\frac{2}{\sqrt{3}} & \frac{2}{\sqrt{3}} & 2 & -1 & 0 \\
\end{array}
\right).
\end{equation}
Reflection with respect to $\vec{a}$-axis leads to
\begin{equation}
(\Lambda^{i1} \ \Lambda^{j2} \ \Lambda^{k3} \ \Lambda^{l4} \ \Lambda^{m5} \ \Lambda^{n6})\Rightarrow 
(-\Lambda^{i4} \ \Lambda^{j3} \ -\Lambda^{k2} \ \Lambda^{l1} \ \Lambda^{m3} \ \Lambda^{n4}),
\end{equation}
so reflection operator is bare reflection $M_\vec{a}^b$ with unitary operator $V_{M_\vec{a}}$,
\begin{equation}
M_\vec{a}=V_{M_\vec{a}} M_\vec{a}^b;\quad V_{M_\vec{a}}=e^{-\frac{\pi}{2}\sum_\vec{r} \left(\phi^{23}_\vec{r} -\phi^{14}_\vec{r} \right) }.
\end{equation}
Similarly, reflection with respect to $(\vec{\hat{a}}+\vec{\hat{b}})$-axis is
\begin{equation}
\begin{split}
M_{\vec{\hat{a}}+\vec{\hat{b}}}&=V_{M_{\vec{\hat{a}}+\vec{\hat{b}}}} M_{\vec{\hat{a}}+\vec{\hat{b}}}^b; \\ V_{M_{\vec{\hat{a}}+\vec{\hat{b}}}}&=e^{-\frac{\pi}{2}\sum_\vec{r} \left(\phi^{12}_\vec{r} +\phi^{34}_\vec{r}+\phi^{45}_\vec{r}+\phi^{56}_\vec{r}-\phi^{36}_\vec{r} \right) }.
\end{split}
\end{equation}

\section{Cubic lattice}\label{sec_4}

In this section we discuss how to apply our $\eta\chi$ formalism to cubic lattice. Comparing to 2d lattice system, the plaquette constraints in 3d lattice systems become more complicated since visually there are many more cycles and many of them are not independent. Since the cubic lattice is also six-coordinated, the operator content is the same as in the triangular lattice. We first consider plaquette constraints and global constraints, with details on how to bosonize fermion-odd operators through the introduction of a ``decorated link'' \cite{po2021symmetric}. We then discuss various symmetries in cubic lattice. We show strategies for finding independent plaquette constraints in Appendix \ref{gen_H}.

\begin{figure}[htpb]
	\begin{tikzpicture}

	\draw[->,  thick] (-3,-0.5)--(-2,-0.5);
	\draw[->,  thick] (-3,-0.5)--(-3,0.5);
	\draw[->,  thick] (-3,-0.5)--(-2.25,0.25);
	\node[above] at (-2,-0.5) {$\hat{\vec{x}}$};
	\node[above] at (-3,0.5) {$\hat{\vec{y}}$};
	\node[above right] at (-2.25,0.25) {$\hat{\vec{z}}$};
	
	\draw[dashed, thick]  (-1,0) -- (1,0);
	\draw[dashed, thick]  (0,-1) -- (0,1);
	\draw[dashed, thick]  (0.9,0.7) -- (-0.9,-0.7);
	\draw[ thick]  (-1,0) -- (0,1);
	\draw[dashed, thick]  (-1,0) -- (0,-1);
	\draw[ thick]  (1,0) -- (0,1);
	\draw[ thick]  (1,0) -- (0,-1);
	\draw[dashed, thick] (-1,0)--(0.9,0.7);
	\draw[ thick] (-1,0)--(-0.9,-0.7); 
	\draw[ thick] (1,0)--(-0.9,-0.7); 
	\draw[ thick] (1,0)--(0.9,0.7); 
	\draw[ thick] (0.9,0.7)--(0,1);
	\draw[ dashed, thick] (0.9,0.7)--(0,-1);
	\draw[thick] (0,-1)--(-0.9,-0.7);
	\draw[thick] (-0.9,-0.7)--(0,1);
	
	\draw[->, double] (1.5,0)--(1.25,0);
	\draw[double] (1.25,0)--(1,0);
	\draw[->, double] (-1,0)--(-1.3,0);
	\draw[double] (-1.3,0)--(-1.5,0);
	\draw[->, double] (0,1.5)--(0,1.25);
	\draw[double] (0, 1.25)--(0,1);
	\draw[->, double] (0, -1)--(0, -1.3);
	\draw[double] (0,-1.3)--(0,-1.5);
	\draw[->, double] (1.3,1.01)--(1.1,0.855);
	\draw[double] (1.1, 0.855)--(0.9,0.7);
	\draw[double] (-1.1,-0.855) -- (-1.3,-1.01);
	\draw[->,double] (-0.9,-0.7) -- (-1.1, -0.855);
	
	\draw [fill] (-1,0) circle [radius=.07];
	\draw [fill] (1,0) circle [radius=.07];
	\draw [fill] (0,-1) circle [radius=.07];
	\draw [fill] (0, 1) circle [radius=.07];
	\draw [fill] (0.9, 0.7) circle [radius=.07];
	\draw [fill] (-0.9, -0.7) circle [radius=.07];
	
	\node[above] at (-1,0+0.1) {3};
	\node[above] at (1,0+0.1) {4};
	\node[left] at (0-0.1,1) {6};
	\node[right] at (0+0.1,-1) {5};
	\node[above left] at(-0.9, -0.7) {1};
	\node[above left] at(0.9, 0.7) {2};

	\end{tikzpicture}
	\captionof{figure}{Majorana fermions attached to a site in cubic lattice with labels. The double lines with arrows represent the sign of mapping $i\ga^i_\vec{r}\ga^j_\vec{r'}$ to products of $\Lambda$s.}
	\label{fig cubic site}	
\end{figure}
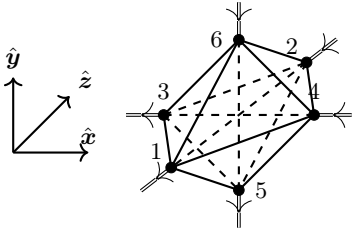

\subsection{Qubit representation}
All operators $\Theta^{ij}_\vec{r},\Lambda^{ij}_\vec{r}$ and $\Phi^{ij}_\vec{r}$ are the same as triangular case. Differences appear in the plaquette constraints. In cubic lattice there are three types of plaquettes, we can call them $xy$-plaquette, $yz$-plaquette and $xz$-plaquette. 
\begin{equation}
\begin{split}
&xy:\quad
\hat{C}_{xy,\vec{r}}\equiv\Phi^{24}_\vec{r}\Phi^{32}_{\vec{r}+\vec{x}}\Phi^{13}_{\vec{r}+\vec{x}+\vec{y}}\Phi^{41}_{\vec{r}+\vec{y}}\ceq-1; \\
&yz:\quad \hat{C}_{yz,\vec{r}}\equiv\Phi^{62}_\vec{r}\Phi^{16}_{\vec{r}+\vec{y}}\Phi^{51}_{\vec{r}+\vec{y}+\vec{z}}\Phi^{25}_{\vec{r}+\vec{z}}\ceq-1; \\
&xz:\quad
\hat{C}_{xz,\vec{r}}=\Phi^{64}_\vec{r}\Phi^{36}_{\vec{r}+\vec{x}}\Phi^{53}_{\vec{r}+\vec{x}+\vec{z}}\Phi^{45}_{\vec{r}+\vec{z}}\ceq-1.
\end{split}
\end{equation} 

These three constraints are not fully independent. To see this, we first assume the cubic lattice has open boundary conditions at least for one direction, for example, $\vec{y}$-direction. Then consider a cube. Applying constraint equations to the pair of $xy$-surfaces, and the pair of $yz$-surfaces, and multiply these equations, we get a product of the pair of $xz$-constraints. We can pick up cubes contiguously along $\vec{y}$-direction, so $xz$-interfaces of these cubes cancel with each other, leaving finally
\begin{equation}
\hat{C}_{xz,\vec{r}}\hat{C}_{xz, \vec{L}_y}=1,
\end{equation}
where $\vec{r}=(r_x, r_y, r_z), \vec{L}_y=(r_x, L_y, r_z)$, i.e. $\vec{L}_y$ is the projection of $\vec{y}$ to the boundary of $\vec{y}$-direction. As long as we fix the boundary $xz$-constraints, in the bulk all $xz$-constraints are automatically true from the other two types of constraints. If the system is periodic along all three directions, i.e., on a 3-torus manifold, then we may choose a reference $xz$-plane, and it is enough to fix the constraints on this $xz$-plane rather than imposing three constraints for each cube.

Graphically we show these constraints as follows:

\begin{equation}
	\begin{tabular}{m{4cm} m{0.5cm} m{3cm} m{0.5cm}}
			\begin{tikzpicture}
			
			\draw[black, thick] (0,0)--(1,0)--(1.7,0.7)--(0.7,0.7)--cycle;

			\filldraw[black] (0,0) circle (2pt);
			\node[left] at (0,0){$\sups{X}{3}\sups{Y}{4}$};
			\node[right] at (1,0){$\sups{X}{3}\sups{X}{4}$};
			\node[above] at (0.3,0.7) {$\sups{Y}{3}\sups{Y}{4}$};
			\node[above ] at(2.1,0.7){$\sups{Y}{3}\sups{X}{4}$};
			\node at(0.9,0.3) {$xy$};

			\end{tikzpicture}
		& 
		\begin{center}
			=
		\end{center}
	    &
		\begin{tikzpicture}
		\node at(-0.5, 0.5) {$(-1)\times$};
		\draw[black, thick] (0,0)--(1,0)--(1.7,0.7)--(0.7,0.7)--cycle;

		\filldraw[black] (0,0) circle (2pt);
		\node at(0.9,0.3) {$xy$};

		\end{tikzpicture}  
		& \quad \\
		
		\begin{tikzpicture}

		\draw[black, thick] (0,0)--(0.8,0.4)--(0.8,1.6)--(0,1.2)--cycle;

		\filldraw[black] (0,0) circle (2pt);
		\node[left] at (0,0){$\sups{Y}{3}$};
		\node[right] at (0.8,0.4){$\sups{X}{3}$};
		\node[above right] at (0.8,1.6){$\sups{Y}{3}\sups{Z}{4}$};
		\node[left] at(0,1.2){$\sups{X}{3}\sups{Z}{4}$};
		\node at(0.4,0.8) {$yz$};

		\end{tikzpicture}
		& 
		\begin{center}
			=
		\end{center}
		&
		\begin{tikzpicture}
		\draw[black, thick] (0,0)--(0.8,0.4)--(0.8,1.6)--(0,1.2)--cycle;

		\filldraw[black] (0,0) circle (2pt);

		\node[left] at (0,0) {$\sups{Z}{1}\sups{Z}{2}$}; 
		\node[right] at(0.8,0.4) {$\sups{Z}{1}\sups{Z}{2}$};
		\node at(0.4,0.8) {$yz$};
		\node at(-0.8,0.8) {$(-1)\times$};
		\end{tikzpicture} 
		& \quad \\
		
		\begin{tikzpicture}
		\draw[black, thick] (0,0)--(1,0)--(1,1)--(0,1)--cycle;
		\filldraw[black] (0,0) circle (2pt);
		\node[left] at (0,0) {$\sups{Z}{3}\sups{Y}{4}$};
		\node[right] at (1,0){$\sups{Z}{3}\sups{X}{4}$};
		\node[right] at (1,1) {$\sups{Y}{4}$};
		\node[left] at(0,1){$\sups{X}{4}$};
		\node at(0.5,0.5) {$xz$};
		\end{tikzpicture}
		& 
		\begin{center}
			=
		\end{center}
		&
		\begin{tikzpicture}
		\draw[black, thick] (0,0)--(1,0)--(1,1)--(0,1)--cycle;
		\filldraw[black] (0,0) circle (2pt);
		\node[left] at (0,0) {$\sups{Z}{1}\sups{Z}{2}$}; 
		\node[right] at(1,0) {$\sups{Z}{1}\sups{Z}{2}$};
		\node at(0.5,0.5) {$xz$};
		\end{tikzpicture} & \quad
	\end{tabular}
\end{equation}

We remark that although these three types of constraints are not independent, it is advantageous to keep track of all of them so as to simplify computing in real applications. We also notice that although we have used parton parity projection, the final forms of plaquette constraints are independent of which subspace of $\Gamma_{\vec{r}}$ the states are projected to.

\subsection{Wilson loops and fermion-odd operators}\label{fermion-odd_cub}
If the system is on a three-dimensional torus, i.e. having three periodic boundary conditions, then we have three Wilson loop constraints. For instance, the fermion loop along $x$-direction gives the identity:
\begin{equation}
(i\gamma_\vec{r}^3\gamma_\vec{r}^4)(i\gamma_\vec{r}^4\gamma_{\vec{r}+\vec{x}}^3)\cdots(i\gamma_{\vec{r}-\vec{x}}^4\gamma_\vec{r}^3)=(-1)^{L_x}.
\end{equation} 
Similarly we have constraints for $y$- and $z$- directions. Mapping these to bosonic operators we obtain:
\begin{equation}
\begin{split}
W_x(\vec{r})&\equiv \prod_{n=0}^{L_x-1}\Phi^{12}_{\vec{r}+n\vec{x}} \ceq -1. \\
W_y(\vec{r})&\equiv \prod_{n=0}^{L_y-1}\Phi^{34}_{\vec{r}+n\vec{y}} \ceq -1. \\
W_z(\vec{r})&\equiv \prod_{n=0}^{L_z-1}\Phi^{56}_{\vec{r}+n\vec{z}} \ceq -1.
\end{split}
\end{equation}

As we mentioned, the three types of plaquette constraints are not independent, so we have to be careful choosing independent constraints for the whole system. In mathematical language, we should choose a set of generators of the free abelian group $H_1$, which is the first homology group of the graph consisting of vertices and edges of cubic lattice. For a generic connected graph $X$, the first homology group can be found with the notion {\itshape{maximal tree}} \cite{Sunada2013}. A tree is defined as a connected subgraph with  no loops, and a maximal tree $T$ is a tree which contains all vertices of $X$. The number of edges in a maximal tree is $|E(T)|=|V(X)|-1$, where ``$|\cdot|$'' means the number of elements in a set \footnote{This equality comes from the fact that a tree can be obtained by joining $|E(T)|$ disconnected edges for $|E(T)|-1$ times. Each time one vertex is resolved into the tree. So $|V(T)|=2|E(T)|-(|E(T)-1|)=|E(T)|+1$. For a maximal tree $|V(T)|=|V(X)|$ so we get the equality.}. Then every edge which is not contained in the maximal tree will be a generator of homology class in $H_1(X)$. Note different choices of maximal trees will give the same homology group. If we denote the set of edges and vertices of a system as $E(X)$ and $V(X)$, then the first homology group is 
\begin{equation}
H_1(X)\cong \mathbb{Z}^{|E(X)|-|E(T)|}= \mathbb{Z}^{|E(X)|-|V(X)|+1}.
\end{equation}
In particular, for a planar graph, the number $(|E(X)|-|E(T)|)$ is equal to the number of ``holes'' in the graph. For non-planar graphs, like embedding on a closed manifold, the choices of generators will be more complicated and depend on details of the graph.  In our lattice system(2d or 3d),  this counting includes both plaquette constraints and Wilson loop constraints. An explicit choice of generators is presented in Appendix \ref{gen_H}.

Before we proceed, we remark that the discussion above is in general applicable for arbitrary connected graphs in which every vertex has an even coordination number. We can count the degrees of freedom and the number of independent constraints for a give graph $X$. For instance, we have $\eta^i_\vec{r},i\in\{1,2,\cdots, 2m_\vec{r}\}$, $\chi^{i}_{\vec{r}}\in\{1,2,\cdots, n_\vec{r}\}$. Coordination number $n_{\vec{r}}$ is the number of edges linking to site $\vec{r}$. Notice that $|E(X)|=\sum_{\vec{r}}n_\vec{r}/2$, so combining the on-site parton parity projection and cycle constraints we obtain
\begin{equation}\label{dim_Hx}
\begin{split}
dim\mathcal{H}_X &=\frac{\prod_\vec{r}2^{m_\vec{r}} 2^{n_\vec{r}/2} }{2^{|V(X)|} 2^{|E(X)|-|V(X)|+1}} \\
&=\frac{2^{|E(X)|}\prod_\vec{r}2^{m_\vec{r}}  }{2^{|E(X)|+1}} \\
&=\frac{1}{2} \prod_\vec{r}2^{m_\vec{r}},
\end{split}
\end{equation}
where the factor 1/2 shows the feature of half Hilbert space with certain global fermionic parity which we will discuss in following. For graphs with odd coordination numbers it is more subtle since to construct local Hilbert space we should have even numbers of Majorana fermions. We discuss these issues in Sec.\ref{sec_5}.

For the cubic lattice, we define
\begin{equation}
\begin{split}
\hat{P} &\equiv\rho^{L_x L_y L_z}\prod_{\vec{r}, r_x=0}W_{x}(\vec{r}) \prod_{\vec{r}, r_y=0}W_{y}(\vec{r}) \prod_{\vec{r}, r_z=0}W_{z}(\vec{r}) \\
&=\rho^{L_x L_y L_z}\prod_{\vec{r}}\Phi^{12}_{\vec{r}}\Phi^{34}_{\vec{r}}\Phi^{56}_{\vec{r}}\\
&=\prod_{\vec{r}}\Theta^{12}_{\vec{r}}\Theta^{34}_{\vec{r}}\Theta^{56}_{\vec{r}}\\
&\ceq (-1)^{L_x+L_y+L_z}\rho^{L_x L_y L_z}.
\end{split}
\end{equation}
The last line shows that if $L_x, L_y,L_z$ are all odd, the parity is $-\rho$; if one of them is odd, the parity is $-1$; otherwise the parity is $+1$. For odd parity case we cannot describe a fermion-even vacuum state, unless we release some of the constraints above. In general, if we want to describe the whole Hilbert space with different fermion parities, we should expect that some fermion-odd operators are also mapped to bosonic side.

In Ref. \onlinecite{po2021symmetric}, this is done by introducing a ``decorated link'' to the square lattice. The basic idea is that in our mapping between fermionic side and bosonic side, there is actually some artificial choice of corresponding signs, i.e. the algebra will be the same for $\pm i\gamma^i_\vec{r} \gamma^j_\vec{r'}\rightarrow \Lambda^{ii}_\vec{r} \Lambda^{jj}_\vec{r'}$ for neighboring sites $\vec{r},\vec{r'}$. By reversing the direction of one of the mapping arrows, four plaquette constraints(living on faces attaching to the edge $\vec{r}\vec{r'}$) and one Wilson loop constraint will have their signs reversed. In practice, we consider
\begin{equation}
	i\gamma^{4}_{-\vec{x}}\gamma^3_\vec{0} \rightarrow -\Lambda^{44}_{-\vec{x}} \Lambda^{33}_\vec{0}.
\end{equation} 
Under this mapping, the modified Wilson loop constraint is now $W_x(n\vec{x})=+1, 0\le n\le L_x-1$, and signs of four plaquette constraints are also reversed.

Consider $\Lambda_{\vec{0}}^{33}$, it is anticommuting with $\Theta^{i3}_\vec{0}, i\neq3$, and commuting with the other $\Theta^{ij}_\vec{0}$s. $\Lambda_\vec{0}^{33}$ also commutes with the decorated link $-\Lambda^{44}_{-\vec{x}}\Lambda^{33}_\vec{0}$. We remind that $\Theta^{i3}_\vec{0}\leftrightarrow i\gamma^{i}_\vec{0}\gamma^3_\vec{0}$. It is clear that $\Lambda_\vec{0}^{33}$ has the same commutation relations with fermion-odd operator $\gamma^{3}_\vec{0}$. Based on this identification, we can map arbitrary Majorana operators in the lattice to bosonic side, thus all fermion-odd operators, by a few number of $\Phi^{ij}$s connecting intermediate sites. For instance, $\gamma^2_{\vec{x}}$ can be expressed as follows:
\begin{equation}
\begin{split}
\gamma^2_{\vec{x}} &=i\gamma^3_\vec{0}\cdot (i \gamma^3_\vec{0} \ga^4_\vec{0}) (i\ga^4_\vec{0} \ga^3_\vec{x}) (i \ga^3_\vec{x} \ga^2_\vec{x}) \\
&\rightarrow i\Lambda^{33}_\vec{0} \Theta^{34}_\vec{0} \Lambda^{44}_\vec{0}\Lambda^{33}_\vec{x} \Theta^{32}_\vec{x} =-i\Phi^{34}_\vec{0}\Lambda^{23}_\vec{x}.\\
\end{split}
\end{equation}

In defining map of fermion-odd operators we have chosen a link to modify the mapping, and one may ask whether the choice is special, since the lattice is homogeneous and a decorated link appears superficially like a defect. Actually, different positions of decorated link can be related by a unitary operator $\hat{\mathcal{M}}(\Phi)$, 
\begin{equation}
\hat{\mathcal{M}}(\Phi)=\frac{1+\hat{P}}{2}+\frac{1-\hat{P}}{2}\Phi,
\end{equation}
where $\Phi$ represents a  Jordan-Wigner string of $\Phi^{ij}_\vec{r}$s. $\hat{P}$ commutes with an arbitrary Jordan-Wigner string $\Phi$. Meanwhile, an operator in the form of products of $\Phi^{ij}_\vec{r}$ and $\Lambda^{ij}_\vec{r}$ will commute or anti-commute with $\hat{P}$ and Jordan-Wigner string $\Phi$. 

We briefly show $\hat M$ moves the decorated link in the following example and refer to Ref. \onlinecite{po2021symmetric} for details.

Consider $\gamma^2_{\vec{x}}$ again, we can take $\Phi=-\Phi^{34}_0\Phi^{32}_\vec{x}$, then one can readily find:
\begin{equation}
\hat{\mathcal{M}}(\Phi) \left(i\Phi^{34}_\vec{0}\Lambda^{23}_\vec{x}\right) \hat{\mathcal{M}}(\Phi)^\dagger=\Lambda^{22}_{\vec{x}}.
\end{equation}
This means after unitary transformation of $\hat{\mathcal{M}}(\Phi)$, $\gamma^2_\vec{x}$ is mapped to $\Lambda^{22}_{\vec{x}}$. Further one can find:
\begin{equation}
\begin{split}
\hat{\mathcal{M}}(\Phi) \left( -\Lambda^{44}_{-\vec{x}}\Lambda^{33}_\vec{0}\right) \hat{\mathcal{M}}(\Phi)^{\dagger} &=\Lambda^{44}_{-\vec{x}}\Lambda^{33}_\vec{0}; \\
\hat{\mathcal{M}}(\Phi) \left( \Lambda^{22}_{\vec{x}}\Lambda^{11}_{\vec{x}+\vec{y}}\right) \hat{\mathcal{M}}(\Phi)^{\dagger} &=-\Lambda^{22}_{\vec{x}}\Lambda^{11}_{\vec{x}+\vec{y}}, \\
\end{split}
\end{equation}
which means the decorated link is moved to $i\ga^{2}_\vec{x}\ga^1_{\vec{x}+\vec{y}}\rightarrow -\Lambda^{22}_{\vec{x}}\Lambda^{11}_{\vec{x}+\vec{y}}$.

\subsection{Symmetries}
In this section we discuss how spatial symmetries of the cubic lattice are represented on the bosonic side.

First we consider translation symmetry $T_\vec{R}: \mathcal{O}_\vec{r}\mapsto \mathcal{O}_{\vec{r}+\vec{R}}$. For odd fermion parity case we have to move the decorated link simultaneously. Similar to discussions in last section, we can use $\hat{\mathcal{M}}(\Phi)$ to move decorated link along three axes, 
\begin{subequations}
\begin{equation}
T_\vec{x}=\hat{\mathcal{M}}(\Phi_\vec{0}^{34})T_\vec{x}^b.
\end{equation}
\begin{equation}
T_\vec{y}=\hat{\mathcal{M}}(\Phi_\vec{0}^{32}\Phi_\vec{y}^{13})T_\vec{y}^b.
\end{equation}
\begin{equation}
T_\vec{z}=\hat{\mathcal{M}}(\Phi_\vec{0}^{36}\Phi_\vec{z}^{53})T_\vec{z}^b.
\end{equation}
\end{subequations} 

Next, consider reflection with respect to $yz$-plane. The bare reflection $M_x^b$ flips the sign of $x$-coordinates for all sites. Meanwhile the auxiliary Majoranas should be reflected as follows,
\begin{equation}
(\Lambda^{i1} \ \Lambda^{j2} \ \Lambda^{k3} \ \Lambda^{l4} \ \Lambda^{m5} \ \Lambda^{n6})\Rightarrow 
(\Lambda^{i1} \ \Lambda^{j2} \ -\Lambda^{k4} \ \Lambda^{l3} \ \Lambda^{m5} \ \Lambda^{n6}).
\end{equation}
For odd fermion parity case, the decorated link is not moved during reflection. So combining internal unitary transformation and bare reflection, the full reflection can be achieved with
\begin{equation}
M_x=V_{M_x}M_x^b;\quad V_{M_x}=e^{-i\frac{\pi}{2}\sum_{\vec{r}}\phi^{34}_\vec{r}}.
\end{equation}

Similarly reflections with respect to $zx$-plane and $xy$-plane can be achieved with
\begin{subequations}
\begin{equation}
M_y=V_{M_y}M_y^b;\quad V_{M_y}=e^{-i\frac{\pi}{2}\sum_{\vec{r}}\phi^{12}_\vec{r}}.
\end{equation}
\begin{equation}
M_z=V_{M_z}M_z^b;\quad V_{M_z}=e^{-i\frac{\pi}{2}\sum_{\vec{r}}\phi^{56}_\vec{r}}.
\end{equation}
\end{subequations}

For rotations of cubic lattice, there are three 4-fold axes, four 3-fold axes and six 2-fold axes. The rotation around $z$-axis  is similar to the discussion in square lattice case Eq. \ref{sq_rotation}, where we just need to composite $C_4$ with an extra $\hat{\mathcal{M}}(\Phi^{31})$ so as to move the decorated link. So here we focus on other axes. Details of transformation matrices of all these axes are collected in Appendix \ref{rot_mat}. Here we first consider $180^\circ$ rotation around 2-axis [110]. It is a bare rotation $C_{2,[110]}^b$ accompanied by a internal unitary which transforms
\begin{equation}\begin{split}
(\Lambda^{i1} \ \Lambda^{j2} \ \Lambda^{k3} \ \Lambda^{l4} \ &\Lambda^{m5} \ \Lambda^{n6})\Rightarrow \\
&(\Lambda^{i3} \ \Lambda^{j4} \ \Lambda^{k1} \ \Lambda^{l2} \ -\Lambda^{m6} \ \Lambda^{n5}).
\end{split}\end{equation}
The full rotation is 
\begin{equation}
\begin{split}
C_{2,[110]} &=\hat{\mathcal{M}}(\Phi^{31}_\vec{r}) V_{C_{2,[110]}}C_{2,[110]}^b; \\
 V_{C_{2,[110]}} &=e^{-i\frac{\pi}{2}\sum_{\vec{r}}\left(\phi^{12}_\vec{r}+\phi^{23}_\vec{r}+\phi^{34}_\vec{r}-\phi^{14}_\vec{r}-\phi^{56}_\vec{r}\right)}.
 \end{split}
\end{equation}

For $120^\circ$ rotation around 3-axis [111], the internal unitary transforms as
\begin{equation}
(\Lambda^{i1} \ \Lambda^{j2} \ \Lambda^{k3} \ \Lambda^{l4} \ \Lambda^{m5} \ \Lambda^{n6})\Rightarrow 
(\Lambda^{i5} \ \Lambda^{j6} \ \Lambda^{k1} \ \Lambda^{l2} \ \Lambda^{m3} \ \Lambda^{n4}).
\end{equation}
The full rotation is 
\begin{equation}
\begin{split}
C_{3,[111]} &=\hat{\mathcal{M}}(\Phi^{31}_\vec{r})  V_{C_{3,[111]}}C_{3,[111]}^b; \\
 V_{C_{3,[111]}} &=e^{-i\frac{2\pi}{3}\sum_{\vec{r}}
 	\frac{1}{\sqrt{3}}\left(\phi^{15}_\vec{r}+\phi^{26}_\vec{r}-\phi^{13}_\vec{r}-\phi^{24}_\vec{r}-\phi^{35}_\vec{r}-\phi^{46}_\vec{r}\right)}.
\end{split}
\end{equation}

\section{Discussions}\label{sec_5}
In this work, we discuss how symmetries could stay manifest on the bosonic side under a higher-dimensional Jordan-Wigner transformation, similar in spirit to the approach in Ref.\ \onlinecite{po2021symmetric}. Our approach relies on the introduction of auxiliary Majorana fermions with clear quantum numbers, and we discuss the case of the square, triangular and cubic lattices as explicit examples. As described, however, our formalism applies only to even-coordinated lattices. In the following, we describe how sites with odd coordination number could be treated, and also draw connections to earlier works related to the design of exactly solved spin liquid models.

First, we address the issue of lattices with odd coordination numbers. As discussed in Sec.\ \ref{sec:graph_count}, by counting the constraints it can be seen that $\eta\chi$ construction is in principle suitable for all kinds of graphs, at least in terms of the Hilbert space dimension. But if some vertices in the graph have odd coordination numbers, then our approach requires introducing an odd number of $\chi$'s on the site, and so we do not have a valid on-site Hilbert space unless the number of $\eta$'s is also odd. 

Given the number of $\eta$ Majorana fermions is fixed by the physical problem of interest, the $\eta\chi$ approach as we discussed is applicable to lattices with oddly coordinated sites only if we start with an odd number of Majorana fermions on such sites. In contrast, most models of interest in condensed matter physics are defined using complex fermions (at least those describing electrons hopping on a lattice), and so the number of Majorana fermions is always even on each site. One possible resolution to this dilemma is to recognize that the graph defining the operator content, and hence the effective coordination number of a site (more accurately, its degree as a vertex on the graph) is not as rigid as it may seem. One could add additional edges to the graph while maintaing the symmetries, such that all sites effectively become even-coordinated.
For instance, consider the trivalent honeycomb lattice. Each site has $3$ nearest neighbors, $6$ second nearest neighbors, and $3$ third nearest neighbors. Therefore, by including also links between third nearest neighbors each site becomes six-coordinated, and we could proceed with the $\eta \chi$ construction without spoiling any spatial symmetries. A trade off, however, is that we introduce more degrees of freedom on the bosonic side, and that there are more loops and hence constraints.

The present $\eta\chi$ formalism can also be related to quantum spin liquid models. Some well-known models like Kitaev's honeycomb model \cite{KITAEV20062} have Majorana fermion representations, and our $\eta\chi$ formalism can also be used to obtain such models in a natural way.  We present two examples here, one is Kitaev honeycomb model, and the other is Ryu's diamond model \cite{PhysRevB.79.075124}. One important feature here is that the fermionic system is taken to be emergent instead of physical, and as such there is no restriction on the number of $\eta$ Majorana fermions per site. It will be natural to leverage such freedom and consider an odd (even) number of $\eta$ fermions on sites with odd (even) coordination numbers.

Consider a honeycomb lattice with one Majorana fermion mode per site. In the $\eta \chi$ formalism $\Lambda^{ij}$ is then a $1\times 3$ matrix, i.e., there are one $\eta$ and three $\chi$'s per site. Nearest neighbor fermion bilinear $i \gamma_{\vec r} \gamma_{\vec r+\vec e_j}$ then gets mapped to $\Lambda^{1j}_{\vec r} \Lambda^{1j}_{\vec r + \vec e_j}$ for $j=1,2,3$ denoting the three neighbor links. Since any two operators among $\{ \Lambda^{11}_{\vec r}, \Lambda^{12}_{\vec r} , \Lambda^{13}_{\vec r}\}$ should anti-commute, and that the site-Hilbert space is two-dimensional (four auxiliary Majorana subjected to a partiy constraint), we naturally get back Kitaev's honeycomb model \cite{KITAEV20062}. Similar constructions with two $\eta$'s and four $\chi$'s per site on the square \cite{PhysRevB.85.155119} and diamond \cite{PhysRevB.79.075124}  lattices would also lead to exactly solved spin liquid models, as we demonstrate in details in Appendix \ref{sec:SL}. With the same logic, it is possible to construct spin liquid models on more general lattices, and we leave this as an interesting future direction.

~\\
\noindent{\em Note Added---}While finishing this work, a related paper appeared \cite{bochniak2021bosonization}. Part of the present work overlaps with Ref.\ \onlinecite{bochniak2021bosonization} in that both discuss how existing methods for bosonizing spinless fermions can be naturally generalize to cover multiple fermion flavors per site.

\appendix{}

\section{Relations between SO(8) representations}\label{irreps}

In our $\eta\chi$ formalism of square lattice,  the qubit representation forms a spinor representation for $SO(8)$, or say, Spin(8). In Ref. \onlinecite{po2021symmetric} the qubit representation is also constructed using auxiliary Majorana fermions. We show that the two ways are physically equivalent. Mathematically it is complicated to figure out a transformation of Clifford basis $\eta,\chi$, but we can take a simpler way, to look for local unitary transformations of qubit representations of operators, such that $\Lambda$'s can be mapped to their expressions in Ref. \onlinecite{po2021symmetric} correspondingly.

We first review some basic facts of SO(8) group \cite{georgi2018lie}. It is a simple Lie group of rank 4. In terms of orthonormal basis vector $\vec{e}^i$ the four simple roots are $\vec{\alpha}^1=\vec{e}^1-\vec{e}^2, \vec{\alpha}^2=\vec{e}^2-\vec{e}^3, \vec{\alpha}^3=\vec{e}^3-\vec{e^4}) $ and $\vec{\alpha}^4=\vec{e}^3+\vec{e^4}$. Its fundamental weights in the same basis are 
\begin{equation}
	\begin{split}
\mu^1 &=(1, 0, 0, 0); \mu^2 =(1, 1, 0, 0); \\
\mu^3 &=\frac{1}{2}(1, 1, 1, -1); \mu^4 =\frac{1}{2}(1, 1, 1, 1).
	\end{split}
\end{equation}
All inequivalent irreducible representations can be constructed from highest weight states $\ket{\mu}, \mu=m^i\mu^i$ with $m^i\in\mathbb{N}$. Fundamental representation $\ket{\mu^1}$ is vector representation, denoted as $\vec{8}_v$. $\ket{\mu^3}$ and $\ket{\mu^4}$ are two spinor representations corresponding to different parity, denoted as $\vec{8}_{s^-}$ and $\vec{8}_{s^+}$ respectively.

In $\eta\chi$ formalism, the generators $\vec{H}=(H_1, H_2, H_3, H_4)$ of Cartan subalgebra are chosen from $\theta^{ij}$ and $\phi^{ij}$ in Eq. \ref{comm}(for convenience we replace $n_{a,b,d,g}$ by $n_{1,2,3,4}$ respectively).
\begin{equation}\label{so8_gen_etachi}
\begin{split}
	H_1&=-\theta^{12}=\frac{1}{2}(2n_1-1); H_2=-\theta^{34}=\frac{1}{2}(2n_2-1);\\
	H_3&=\phi^{12}=\frac{1}{2}(2n_3-1); H_4=\phi^{34}=\frac{1}{2}(2n_4-1);\\
\end{split}
\end{equation}
Then irrep $\vec{8}_{s^-}$ has highest weight state $\ket{1110}$.

\begin{figure}[htbp]
	\begin{tabular}{m{4cm} m{4cm}}
		\begin{center}
\begin{tikzpicture}
\node[shape=rectangle,draw=black] (A) at (0,3) {$\ket{1110}$};
\node[shape=rectangle,draw=black] (B) at (0,2) {$\ket{1101}$};
\node[shape=rectangle,draw=black] (C) at (0,1) {$\ket{1011}$};
\node[shape=rectangle,draw=black] (D) at (-1,0) {$\ket{0111}$};
\node[shape=rectangle,draw=black] (E) at (1, 0) {$\ket{1000}$};
\node[shape=rectangle,draw=black] (F) at (0,-1) {$\ket{0100}$} ;
\node[shape=rectangle,draw=black] (G) at (0,-2) {$\ket{0010}$} ;
\node[shape=rectangle,draw=black] (H) at (0,-3) {$\ket{0001}$} ;

\path [->] (A) edge  (B);
\path [->](B) edge  (C);
\path [->](C) edge  (D);
\path [->](C) edge (E);
\path [->](D) edge  (F);
\path [->](E) edge  (F);
\path [->](F) edge (G);
\path [->](G) edge (H);

\end{tikzpicture}
\end{center}
&
\begin{center}
\begin{tikzpicture}

\node[shape=rectangle,draw=black] (A) at (0,3) {$\ket{1000}'$};
\node[shape=rectangle,draw=black] (B) at (0,2) {$\ket{1011}'$};
\node[shape=rectangle,draw=black] (C) at (0,1) {$\ket{0010}'$};
\node[shape=rectangle,draw=black] (D) at (-1,0) {$\ket{0001}'$};
\node[shape=rectangle,draw=black] (E) at (1, 0) {$\ket{1110}'$};
\node[shape=rectangle,draw=black] (F) at (0,-1) {$\ket{1101}'$} ;
\node[shape=rectangle,draw=black] (G) at (0,-2) {$\ket{0100}'$} ;
\node[shape=rectangle,draw=black] (H) at (0,-3) {$\ket{0111}'$} ;

\path [->] (A) edge  (B);
\path [->](B) edge  (C);
\path [->](C) edge  (D);
\path [->](C) edge (E);
\path [->](D) edge  (F);
\path [->](E) edge  (F);
\path [->](F) edge (G);
\path [->](G) edge (H);

\end{tikzpicture}
\end{center}
\end{tabular}
\caption{Weight diagram of two irreps.}
\label{irrep_weights}
\end{figure}
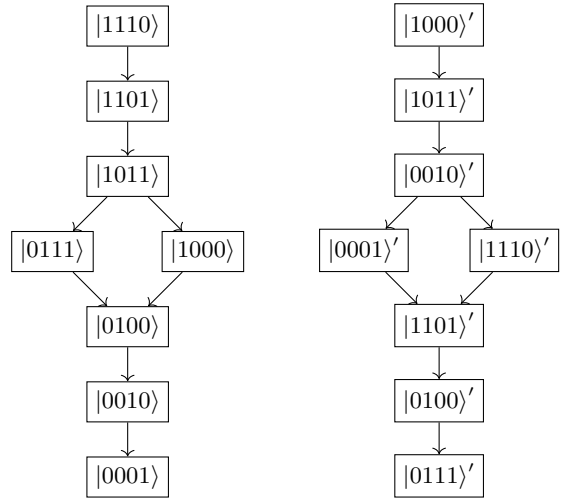

Although it is not obvious how to find a local unitary operator by $\eta$'s and $\chi$'s such that our formalism here can be turned into ones of spin charge separation in Ref.\ \onlinecite{po2021symmetric}, we can first consider local unitary transformations between qubit representations. We look for a unitary matrix $U$ such that
\begin{equation}
    U\Lambda^{ij}U^{-1}=\Lambda^{ij}_{sc},
\end{equation}
where $\Lambda_{sc}$ means the corresponding expression in Eq. (112) of Ref. \onlinecite{po2021symmetric}. When $\rho=-1$, this $8\times 8$ matrix $U$ turns out to be
\begin{equation}
U=\left(
    \begin{array}{cccc}
        0 & 0 & 0 & X  \\
        0 & 0 & -iX & 0 \\
        X & 0 & 0 & 0 \\
        0 & -iX & 0 & 0
    \end{array}
\right),
\end{equation}
where $X$ is $2\times2$ Pauli matrix, and each entry denotes a $2\times2$ block. Under this transformation, the generators in Eq. \ref{so8_gen_etachi} are mapped to another form. We can readily check that the states after being mapped by $U$ still form an $\vec{8}_{s^-}$ irrep. More explicitly, for a state with weight $\ket{\mu^i}$ in $\eta\chi$ quib representation, if we operate transformed Cartan generators on these states,
\begin{equation}
    UH_\alpha U^{-1}\cdot U\ket{\mu^i}=\mu^i_\alpha U\ket{\mu^i}, \alpha\in\{1,2,3,4\},
\end{equation}
which means $\ket{\mu^i}$ still has weight $\mu^i$. We have summarized the weight diagram of states in $\eta\chi$ qubit representation and the transformed qubit representation. Notice that in Ref. \cite{po2021symmetric} spins representations are defined from particle number representations in a different way from ours.

After unitary transformation the irrep $\vec{8}_{s^-}$ has highest weight state $\ket{1000}'$(we use $'$ to distinguish states in the sense of  spin-charge separation). Notice in Ref. \onlinecite{po2021symmetric} the particle number is $n=1+n_c-n_h$, and z-component spin is $S^z=\frac{1}{2}(n_u-n_d)$. So $\ket{1000}'$ represents a state $n=2, S^z=0$, which is the same as $\ket{1110}$ in $\eta\chi$ formalism(if we ignore the auxiliary degrees of freedom). Similarly one may compare all descendants in the two irreps as shown in Fig. \ref{irrep_weights}.

\section{Generators of $H_1(X,\mathbb{Z})$ of connected graphs}\label{gen_H}
\begin{figure}
	\begin{tabular}{m{4cm} m{2cm}}

\begin{tikzpicture}[global scale=1]
\path[coordinate] (0,0)  coordinate(A)
--( 1cm,1cm) coordinate(B)
--( 2cm,1.3cm) coordinate(C)
--( 2.7cm,0.8cm) coordinate(D)
--(3.3cm,-0.2cm) coordinate(E)
--(1.6cm,-0.9cm) coordinate(F)
--(1.2cm,-0.3cm) coordinate(G)
--(1.8,0.5) coordinate(M)
--(0.3,-0.6) coordinate(L)
;
\draw (A) -- (B) --(C)--(D)--(E)--(F)--(G)--cycle;
\draw (A) -- (M) --(B);
\draw (C) -- (M) --(D) --(G);
\draw (D) --(F)--(L)--(A);
\draw[ultra thick] (A)--(B)--(C)--(D)--(E);
\draw[ultra thick] (A)--(M);
\draw[ultra thick] (A)--(G);
\draw[ultra thick] (A)--(L)--(F);
\node[left] at (A){A};
\node[left] at (B){B};
\node[above] at (C){C};
\node[right] at (D){D};
\node[below] at (E){E};
\node[below] at (F){F};
\node[right] at (G){G};
\node[left] at (1.6,0.6) {M};
\node[below] at (L){L};
\end{tikzpicture}

&
\begin{center}
\begin{tikzpicture}
\draw [fill] (0,0) circle [radius=.1];
\foreach \x in {0,1}{
	\draw (\x, 0)--(\x, 2);
	\draw (0, \x)--(2, \x);
	\node at (\x, 2) {$\times$};
	\node at (2, \x) {$\times$};
};
\node at (1, 0.5) {$\times$};
\node[left] at (0,2) {$W_y$};
\node[right] at (2,0) {$W_x$};
\draw[ultra thick] (1,0)--(0,0)--(0,1)--(1,1);
\draw[->, thick] (1,2) to [out=180, in=85] (0.5,1.5);
\draw[->, thick] (2,1) to [out=270, in=5] (1.5,0.5);
\draw[->, thick] (1,0.5)--(0.5, 0.5);
\end{tikzpicture}
\end{center}
\end{tabular}
\caption{A planar graph with some vertices and edges. The thick lines consists a maximal tree $T$, and every edge not belonging to $T$ generates a homology class, namely, a 1-cycle. The number of generators is equal to the number of ``holes''(Euler characteristic).}
\label{rand_graph}
\end{figure}
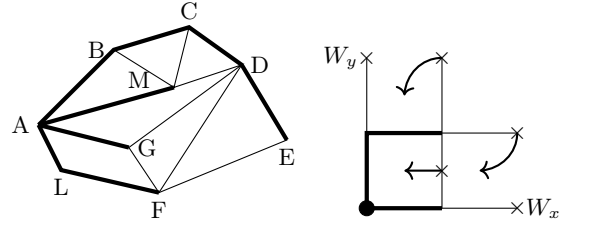

In this appendix we discuss how to find generators of the first homology group for a connected graph. For example, Fig. \ref{rand_graph} is a planar graph consists of 9 vertices and 15 edges. A maximal tree $T$ is emphasized by thick lines. Then for every edge $e\notin T$, adding it to this maximal tree will produce a class of cycle, like adding BM to $T$ will produce a cycle ABM. In the language of bosonization, each cycle of fermionic side is an identity, while mapping to bosonic side gives a constraint.  In Fig. \ref{rand_graph} there are 7 edges not included in $T$ and they generate the whole $H_1{X,\mathbb{Z}}\cong \mathbb{Z}^{7}$.

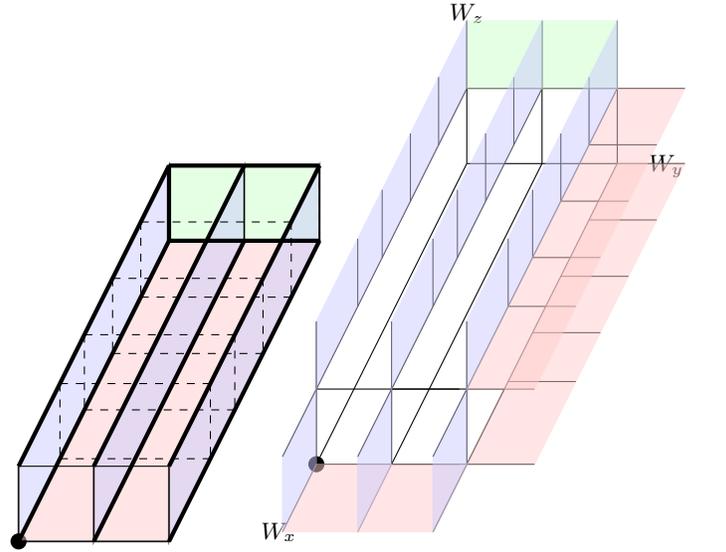
\begin{figure}[htbp]
	\begin{center}
	\begin{tabular}{m{3cm} m{6cm}}
		\begin{center}
\begin{tikzpicture}
\draw [fill] (0,0) circle [radius=.1];

\fill [fill=blue!20,draw=black,thick,opacity=0.5]
(0,0) -- (2,4)--(2,5) --(0,1)--cycle;
\fill [fill=red!20,draw=black,thick,opacity=0.5]
(0,0) -- (2,0)--(4,4) --(2,4)--cycle;
\fill [fill=green!20,draw=black,thick,opacity=0.5]
(2,4) -- (4,4)--(4,5) --(2,5)--cycle;
\fill [fill=blue!20,draw=black,thick,opacity=0.5]
(2,0) -- (4,4)--(4,5) --(2,1)--cycle;
\fill [fill=blue!20,draw=black,thick,opacity=0.5]
(1,0) -- (3,4)--(3,5) --(1,1)--cycle;

\foreach \x in {0,1,2}{
	\draw[ultra thick] (\x,0)--(2+\x, 4);
	\draw[ultra thick] (\x,1)--(2+\x, 5);
	\draw (\x,0)--(\x,1);
	\draw (2+\x,4)--(2+\x, 5);
};
\foreach \x in {0,1}{
	\draw (\x, 0)--(\x+1,0);
	\draw (\x, 1)--(\x+1, 1);
	\draw[ultra thick] (2+\x, 4)--(2+\x+1, 4);
	\draw[ultra thick] (2+\x, 5)--(2+\x+1, 5);
};
\foreach \x in {0.1, 0.75, 1.5, 2.25, 3}{
	\draw[dashed](0.5+0.5*\x,1+\x)--(2.5+0.5*\x,1+\x);
	\draw[dashed](0.5+0.5*\x,2+\x)--(2.5+0.5*\x,2+\x);
	\draw[dashed](2.5+0.5*\x,2+\x)--(2.5+0.5*\x,1+\x);
	\draw[dashed](0.5+0.5*\x,2+\x)--(0.5+0.5*\x,1+\x);
};
	
	\draw[ultra thick] (2, 4)--(2, 5);
	
\end{tikzpicture}
\end{center}
&
\begin{center}
\begin{tikzpicture}
\draw [fill] (0,0) circle [radius=.1];
\foreach \x in {0,1,2}{
	\draw (\x,0)--(2+\x, 4);
	\draw (\x,1)--(2+\x, 5);
	\draw (\x,0)--(\x,1+0.9);
	\draw (2+\x,4)--(2+\x, 5+0.9);
	\draw (\x,0)--(\x-0.45, -0.9);
	\draw (\x,1)--(\x-0.45, 0.1);
};
\foreach \x in {0,1}{
	\draw (\x, 0)--(\x+1+0.9,0);
	\draw (\x, 1)--(\x+1+0.9, 1);
	\draw (2+\x, 4)--(2+\x+1, 4);
	\draw (2+\x, 5)--(2+\x+1, 5);
};
\foreach \x in {0.1, 0.75, 1.5, 2.25, 3}{
	\draw (2.5+0.5*\x,1+\x)--(2.5+0.5*\x+0.9,1+\x);
	\draw (2.5+0.5*\x,2+\x)--(2.5+0.5*\x+0.9,2+\x);
	\draw (2.5+0.5*\x,2+\x)--(2.5+0.5*\x,2+\x+0.9);
	\draw (0.5+0.5*\x,2+\x)--(0.5+0.5*\x,2+\x+0.9);
	\draw (1.5+0.5*\x,2+\x)--(1.5+0.5*\x,2+\x+0.9);
};

	\node at(-0.5, -0.9) {$W_x$};
	\node[above left] at(5, 3.7) {$W_y$};
	\node at(2, 6) {$W_z$};
	
\foreach \x in {0, 1}{	
	\fill [fill=green!20,draw=green!20,thin,opacity=0.5]
	(2+\x,5) -- (3+\x,5)--(3+\x,5.9) --(2+\x,5.9)--cycle;
}

\foreach \x in {0, 1}{	
	\fill [fill=red!20,draw=red!20,thin,opacity=0.5]
	(-0.45+\x,-0.9) -- (0.55+\x,-0.9)--(1+\x,0) --(0+\x,0)--cycle;
}

\foreach \x in{0,1}{
	\fill [fill=red!20, draw=red!20, thin, opacity=0.5]
	(2,\x)--(2.9,\x)--(4.9, \x+4)--(4,\x+4)--cycle;
}

\foreach \x in {0, 1, 2}{	
	\fill [fill=blue!20,draw=blue!20,thin,opacity=0.5]
	(0+\x,1) -- (2+\x,5)--(2+\x,5.9) --(0+\x,1.9)--cycle;
	\fill [fill=blue!20,draw=blue!20,thin,opacity=0.5]
	(-0.45+\x,-0.9) -- (0+\x,0)--(0+\x,1) --(-0.45+\x,0.1)--cycle;
}

\end{tikzpicture}
\end{center}
\end{tabular}
\end{center}

\caption{Plaquette constraints for cubic lattice on three-torus. The left figure shows independent plaquettes colored in cuboid(for vision the upper surface is not exhibited). The right figure shows Wilson loops and extra plaquettes from periodic boudary conditions,}
\label{Cubic_plaquettes}
\end{figure}

For non-planar connected graph, for instance, a lattice with periodic boundary conditions, the counting of generators of $H_1(X,\mathbb{Z})$ is also similar, but the meaning of ``holes'' becomes tricky. In generic situation, we can start from one vertex of the graph and find a maximal tree, count edges not included in the tree, and assign each of them with a cycle. We can start from an example of $2\times2$ square lattice.

In Fig. \ref{rand_graph} the dangling edges with crosses(``crossed edge'') are linking to periodic sites respectively. The maximal tree is stressed with thick lines. The left most crossed edge gives the Wilson loop constraint along $y$-direction, and the bottom crossed edge gives the Wilson loop constraint along $x$-direction. The right edge of the square gives plaquette to its left. The two crossed edges at the top-right corner gives two plaquettes to their left and right.

Now we consider cubic lattice. To make it concrete we set $L_x=6, L_y=3, L_z=2$. We first consider the $(L_x-1)\times(L_y-1)\times(L_z-1)$ cuboid, its maximal tree is chosen to be double layers of ``E'' shape with an extra edge connecting two layers. Using edges in this cuboid we can get plaquettes living on five surfaces of the cuboid. We use different colors to shade the plaquette constraints from those crossed edges, these surfaces includes ($y=0,xz$), ($y=2, xz$), ($x=0, yz$),($z=0, xy$), ($z=1,xy$). Now we turn to those dangling edges. The strategy is as follows. First determine the Wilson loop constraints, and then link each crossed edge to a plaquette on the extension of cuboid surfaces mentioned earlier. This strategy makes Wilson loops move freely in the whole graph, and plaquette constraints are independent and easy to count.

The counting of plaquette is straightforward. Combining the difference contributions, we have 
\[
\begin{split}
\text{\#}\{\text{constraints}\}&=(L_xL_z-1)L_y+(L_y-1)(L_x-1)L_z \\
&+(L_x-1)L_z+(L_yL_z-1)+(L_y-1) +3 \\
&=2L_x L_y L_z+1.
\end{split}
\]
The last line is exactly the same as the exponent of the denominator in Eq. \ref{dim_Hx}.

\section{Rotation matrices and exponential maps}\label{rot_mat}

In this section we list some useful equations of rotation matrices of cubic lattice for all types of axes. As the notation in main text, we denote a $6\times 6$ rotation matrix as $R$, for $\vec{\Lambda}^{T} \Rightarrow \vec{\Lambda}^{T} R$, and  their exponential form $R=e^{-2A}$ where $A$ is a $6\times 6$ anti-symmetric real matrix. We summarize equations of exponential maps so that one can readily write down the corresponding unitary operator of rotation.

For 4-fold axes we just use the $V_{C_4}$ of square lattice to write down similar operators in cubic lattice.
\begin{subequations}

\begin{equation}
V_{[0,0,1]}=e^{-i\frac{\pi}{4}\sum_\vec{r}\left( \phi^{12}_\vec{r} +\phi^{34}_\vec{r}-\sqrt{2} (\phi^{13}_\vec{r}+\phi^{24}_\vec{r}+\phi^{14}_\vec{r}-\phi^{23}_\vec{r}) \right)},
\end{equation}

\begin{equation}
V_{[1,0,0]}=e^{-i\frac{\pi}{4}\sum_\vec{r}\left( \phi^{56}_\vec{r} +\phi^{12}_\vec{r}-\sqrt{2} (\phi^{51}_\vec{r}+\phi^{62}_\vec{r}+\phi^{52}_\vec{r}-\phi^{61}_\vec{r}) \right)},
\end{equation}

\begin{equation}
V_{[0,-1,0]}=e^{-i\frac{\pi}{4}\sum_\vec{r}\left( \phi^{56}_\vec{r} +\phi^{34}_\vec{r}-\sqrt{2} (\phi^{53}_\vec{r}+\phi^{64}_\vec{r}+\phi^{54}_\vec{r}-\phi^{63}_\vec{r}) \right)},
\end{equation}

\end{subequations}

2-fold axes: 
\begin{widetext}
\begin{equation}
\begin{array}{ccc}
[1, 1, 0]:
&
R=\left(
\begin{array}{cccccc}
0 & 0 & 1 & 0 & 0 & 0 \\
0 & 0 & 0 & 1 & 0 & 0 \\
1 & 0 & 0 & 0 & 0 & 0 \\
0 & 1 & 0 & 0 & 0 & 0 \\
0 & 0 & 0 & 0 & 0 & 1 \\
0 & 0 & 0 & 0 & -1 & 0 \\
\end{array}
\right)
&
=\text{exp}[-\frac{\pi}{2}\left(
\begin{array}{cccccc}
0 & 1 & 0 & -1 & 0 & 0 \\
-1 & 0 & 1 & 0 & 0 & 0 \\
0 & -1 & 0 & 1 & 0 & 0 \\
1 & 0 & -1 & 0 & 0 & 0 \\
0 & 0 & 0 & 0 & 0 & -1 \\
0 & 0 & 0 & 0 & 1 & 0 \\
\end{array}
\right)].
\end{array}
\end{equation}

\begin{equation}
\begin{array}{ccc}
[1, -1, 0]:
&
R=\left(
\begin{array}{cccccc}
0 & 0 & 0 & 1 & 0 & 0 \\
0 & 0 & -1 & 0 & 0 & 0 \\
0 & 1 & 0 & 0 & 0 & 0 \\
-1 & 0 & 0 & 0 & 0 & 0 \\
0 & 0 & 0 & 0 & 0 & 1 \\
0 & 0 & 0 & 0 & -1 & 0 \\
\end{array}
\right)
&
=\text{exp}[-\frac{\pi}{2}\left(
\begin{array}{cccccc}
0 & 0 & 0 & -1 & 0 & 0 \\
0 & 0 & 1 & 0 & 0 & 0 \\
0 & -1 & 0 & 0 & 0 & 0 \\
1 & 0 & 0 & 0 & 0 & 0 \\
0 & 0 & 0 & 0 & 0 & -1 \\
0 & 0 & 0 & 0 & 1 & 0 \\
\end{array}
\right)].
\end{array}
\end{equation}

\begin{equation}
\begin{array}{ccc}
[1, 0, 1]:
&
R=\left(
\begin{array}{cccccc}
0 & 0 & 0 & 0 & 1 & 0 \\
0 & 0 & 0 & 0 & 0 & 1 \\
0 & 0 & 0 & -1 & 0 & 0 \\
0 & 0 & 1 & 0 & 0 & 0 \\
1 & 0 & 0 & 0 & 0 & 0 \\
0 & 1 & 0 & 0 & 0 & 0 \\
\end{array}
\right)
&
=\text{exp}[-\frac{\pi}{2}\left(
\begin{array}{cccccc}
0 & 1 & 0 & 0 & 0 & -1 \\
-1 & 0 & 0 & 0 & 1 & 0 \\
0 & 0 & 0 & 1 & 0 & 0 \\
1 & 0 & -1 & 0 & 0 & 0 \\
0 & -1 & 0 & 0 & 0 & 1 \\
1 & 0 & 0 & 0 & -1 & 0 \\
\end{array}
\right)].
\end{array}
\end{equation}

\begin{equation}
\begin{array}{ccc}
[-1, 0, 1]:
&
R=\left(
\begin{array}{cccccc}
0 & 0 & 0 & 0 & 0 & 1 \\
0 & 0 & 0 & 0 & -1 & 0 \\
0 & 0 & 0 & -1 & 0 & 0 \\
0 & 0 & 1 & 0 & 0 & 0 \\
0 & 1 & 0 & 0 & 0 & 0 \\
-1 & 0 & 0 & 0 & 0 & 0 \\
\end{array}
\right)
&
=\text{exp}[-\frac{\pi}{2}\left(
\begin{array}{cccccc}
0 & 0 & 0 & 0 & 0 & -1 \\
0 & 0 & 0 & 0 & 1 & 0 \\
0 & 0 & 0 & 1 & 0 & 0 \\
0 & 0 & -1 & 0 & 0 & 0 \\
0 & -1 & 0 & 0 & 0 & 0 \\
1 & 0 & 0 & 0 & 0 & 0 \\
\end{array}
\right)].
\end{array}
\end{equation}

\begin{equation}
\begin{array}{ccc}
[0, 1, 1]:
&
R=\left(
\begin{array}{cccccc}
0 & -1 & 0 & 0 & 0 & 0 \\
1 & 0 & 0 & 0 & 0 & 0 \\
0 & 0 & 0 & 0 & 1 & 0 \\
0 & 0 & 0 & 0 & 0 & 1 \\
0 & 0 & 1 & 0 & 0 & 0 \\
0 & 0 & 0 & 1 & 0 & 0 \\
\end{array}
\right)
&
=\text{exp}[-\frac{\pi}{2}\left(
\begin{array}{cccccc}
0 & 1 & 0 & 0 & 0 & 0 \\
-1 & 0 & 0 & 0 & 0 & 0 \\
0 & 0 & 0 & 1 & 0 & -1 \\
0 & 0 & -1 & 0 & 1 & 0 \\
0 & 0 & 0 & -1 & 0 & 1 \\
0 & 0 & 1 & 0 & -1 & 0 \\
\end{array}
\right)].
\end{array}
\end{equation}

\begin{equation}
\begin{array}{ccc}
[0, -1, 1]:
&
R=\left(
\begin{array}{cccccc}
0 & -1 & 0 & 0 & 0 & 0 \\
1 & 0 & 0 & 0 & 0 & 0 \\
0 & 0 & 0 & 0 & 0 & 1 \\
0 & 0 & 0 & 0 & -1 & 0 \\
0 & 0 & 0 & 1 & 0 & 0 \\
0 & 0 & -1 & 0 & 0 & 0 \\
\end{array}
\right)
&
=\text{exp}[-\frac{\pi}{2}\left(
\begin{array}{cccccc}
0 & 1 & 0 & 0 & 0 & 0 \\
-1 & 0 & 0 & 0 & 0 & 0 \\
0 & 0 & 0 & 0 & 0 & -1 \\
0 & 0 & 0 & 0 & 1 & 0 \\
0 & 0 & 0 & -1 & 0 & 0 \\
0 & 0 & 1 & 0 & 0 & 0 \\
\end{array}
\right)].
\end{array}
\end{equation}

3-fold axes:
\begin{equation}
\begin{array}{ccc}
[1, 1, 1]:
&
R=\left(
\begin{array}{cccccc}
0 & 0 & 1 & 0 & 0 & 0 \\
0 & 0 & 0 & 1 & 0 & 0 \\
0 & 0 & 0 & 0 & 1 & 0 \\
0 & 0 & 0 & 0 & 0 & 1 \\
1 & 0 & 0 & 0 & 0 & 0 \\
0 & 1 & 0 & 0 & 0 & 0 \\
\end{array}
\right)
&
=\text{exp}[
\frac{2\pi}{3\sqrt{3}}\left(
\begin{array}{cccccc}
0 & 0 & 1 & 0 & -1 & 0 \\
0 & 0 & 0 & 1 & 0 & -1 \\
-1 & 0 & 0 & 0 & 1 & 0 \\
0 & -1 & 0 & 0 & 0 & 1 \\
1 & 0 & -1 & 0 & 0 & 0 \\
0 & 1 & 0 & -1 & 0 & 0 \\
\end{array}
\right)
].
\end{array}
\end{equation}

\begin{equation}
\begin{array}{ccc}
[1, -1, 1]:
&
R=\left(
\begin{array}{cccccc}
0 & 0 & 0 & 0 & 0 & -1 \\
0 & 0 & 0 & 0 & 1 & 0 \\
0 & 1 & 0 & 0 & 0 & 0 \\
-1 & 0 & 0 & 0 & 0 & 0 \\
0 & 0 & 1 & 0 & 0 & 0 \\
0 & 0 & 0 & 1 & 0 & 0 \\
\end{array}
\right)
&
=\text{exp}[
\frac{2\pi}{3\sqrt{3}}\left(
\begin{array}{cccccc}
0 & 0 & 0 & 1 & 0 & -1 \\
0 & 0 & -1 & 0 & 1 & 0 \\
0 & 1 & 0 & 0 & -1 & 0 \\
-1 & 0 & 0 & 0 & 0 & -1 \\
0 & -1 & 1 & 0 & 0 & 0 \\
1 & 0 & 0 & 1 & 0 & 0 \\
\end{array}
\right)
].
\end{array}
\end{equation}

\begin{equation}
\begin{array}{ccc}
[-1, -1, 1]:
&
R=\left(
\begin{array}{cccccc}
0 & 0 & 1 & 0 & 0 & 0 \\
0 & 0 & 0 & 1 & 0 & 0 \\
0 & 0 & 0 & 0 & 0 & -1 \\
0 & 0 & 0 & 0 & 1 & 0 \\
0 & 1 & 0 & 0 & 0 & 0 \\
-1 & 0 & 0 & 0 & 0 & 0 \\
\end{array}
\right)
&
=\text{exp}[
\frac{2\pi}{3\sqrt{3}}\left(
\begin{array}{cccccc}
0 & 0 & 1 & 0 & 0 & 1 \\
0 & 0 & 0 & 1 & -1 & 0 \\
-1 & 0 & 0 & 0 & 0 & -1 \\
0 & -1 & 0 & 0 & 1 & 0 \\
0 & 1 & 0 & -1 & 0 & 0 \\
-1 & 0 & 1 & 0 & 0 & 0 \\
\end{array}
\right)
].
\end{array}
\end{equation}

\begin{equation}
\begin{array}{ccc}
[-1, 1, 1]:
&
R=\left(
\begin{array}{cccccc}
0 & 0 & 0 & 0 & 1 & 0 \\
0 & 0 & 0 & 0 & 0 & 1 \\
0 & 1 & 0 & 0 & 0 & 0 \\
-1 & 0 & 0 & 0 & 0 & 0 \\
0 & 0 & 0 & -1 & 0 & 0 \\
0 & 0 & 1 & 0 & 0 & 0 \\
\end{array}
\right)
&
=\text{exp}[
\frac{2\pi}{3\sqrt{3}}\left(
\begin{array}{cccccc}
0 & 0 & 0 & 1 & 1 & 0 \\
0 & 0 & -1 & 0 & 0 & 1 \\
0 & 1 & 0 & 0 & 0 & -1 \\
-1 & 0 & 0 & 0 & 1 & 0 \\
-1 & 0 & 0 & -1 & 0 & 0 \\
0 & -1 & 1 & 0 & 0 & 0 \\
\end{array}
\right)
].
\end{array}
\end{equation}

\end{widetext}

\section{Connection to spin liquid models \label{sec:SL}}

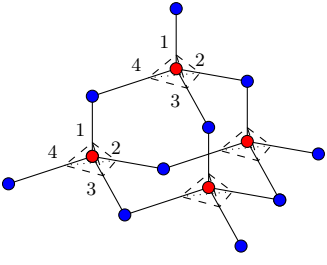
\begin{figure}[htbp]
	
	\begin{center}
	\begin{tikzpicture}[global scale=0.8]
		\path [coordinate] (0,0)  coordinate(A)
		--(345:2) coordinate(B)
		-- ++(50:1) coordinate(C)
		-- ++(170:1.2 ) coordinate(D)
		-- ++(90:1) coordinate(E);
		\draw (A)--(D);
		\draw (B)--(D);
		\draw (C)--(D);
		\draw (D)--(E);
		\draw[fill=red] (D) circle (0.1);
		
		\path [coordinate] (B)
		-- ++(345:2) coordinate(M)
		-- ++(50:1) coordinate(N)
		-- ++(170:1.2 ) coordinate(P)
		-- ++(90:1) coordinate(Q);
		\draw (B)--(P);
		\draw (M)--(P);
		\draw (N)--(P);
		\draw (P)--(Q);
		
		\path [coordinate] (C)
		-- ++(345:2) coordinate(H)
		-- ++(50:1) coordinate(J)
		-- ++(170:1.2 ) coordinate(K)
		-- ++(90:1) coordinate(L);
		\draw (C)--(K);
		\draw (H)--(K);
		\draw (J)--(K);
		\draw (K)--(L);
		
		\path [coordinate] (L)-- ++(170:1.2 ) coordinate(S)
		-- ++(90:1) coordinate(T);
		\draw (E)--(S);
		\draw (Q)--(S);
		\draw (L)--(S);
		\draw (S)--(T);
		
		\draw[fill=red] (D) circle (0.1);
		\draw[fill=red] (P) circle (0.1);
		\draw[fill=red] (K) circle (0.1);
		\draw[fill=red] (S) circle (0.1);
		\draw[fill=blue](A) circle (0.1);
		\draw[fill=blue](B) circle (0.1);
		\draw[fill=blue](C) circle (0.1);
		\draw[fill=blue](E) circle (0.1);
		\draw[fill=blue](M) circle (0.1);
		\draw[fill=blue](N) circle (0.1);
		\draw[fill=blue](Q) circle (0.1);
		\draw[fill=blue](H) circle (0.1);
		\draw[fill=blue](J) circle (0.1);
		\draw[fill=blue](L) circle (0.1);
		\draw[fill=blue](T) circle (0.1);
		
		\path [coordinate] (T)-- ++(270:0.77) coordinate(ak);
		\path [coordinate] (S)-- ++(350:0.4) coordinate(bk)
		-- ++(230:0.33) coordinate(ck)
		-- ++(165:0.667) coordinate(dk);
		\draw[dashed] (ak)--(bk)--(ck)--(dk)--cycle;
		\draw[dashed] (ak)--(ck);
		\draw[dotted] (bk)--(dk);
		\node[above left] at(ak){1};
		\node[above] at(bk){2};
		\node[below left] at(ck){3};
		\node[above left] at(dk){4};
		
		\path [coordinate] (E)-- ++(270:0.77) coordinate(ak);
		\path [coordinate] (D)-- ++(350:0.4) coordinate(bk)
		-- ++(230:0.33) coordinate(ck)
		-- ++(165:0.667) coordinate(dk);
		\draw[dashed] (ak)--(bk)--(ck)--(dk)--cycle;
		\draw[dashed] (ak)--(ck);
		\draw[dotted] (bk)--(dk);
		\node[above left] at(ak){1};
		\node[above] at(bk){2};
		\node[below left] at(ck){3};
		\node[above left] at(dk){4};
		
		\path [coordinate] (Q)-- ++(270:0.77) coordinate(ak);
		\path [coordinate] (P)-- ++(350:0.4) coordinate(bk)
		-- ++(230:0.33) coordinate(ck)
		-- ++(165:0.667) coordinate(dk);
		\draw[dashed] (ak)--(bk)--(ck)--(dk)--cycle;
		\draw[dashed] (ak)--(ck);
		\draw[dotted] (bk)--(dk);
		
		\path [coordinate] (L)-- ++(270:0.77) coordinate(ak);
		\path [coordinate] (K)-- ++(350:0.4) coordinate(bk)
		-- ++(230:0.33) coordinate(ck)
		-- ++(165:0.667) coordinate(dk);
		\draw[dashed] (ak)--(bk)--(ck)--(dk)--cycle;
		\draw[dashed] (ak)--(ck);
		\draw[dotted] (bk)--(dk);
		
	\end{tikzpicture}
	\end{center}

\caption{Ryu's model of diamond lattice. The figure shows two types of sites in diamond lattice with red A type and blue B type. Numbers are labeling the edge to which a pair of $\chi^i$'s are attached. Each site has a tetrahedron. Tetrahedrons of blue sites are not shown in figure.}
\label{fig_models}
\end{figure}

We consider a diamond lattice. There are two sets of sublattices, and each vertex has four types of links to its nearest neighbors. Each vertex can be assigned a tetrahedron, whose four vertices correspond to $\chi^i, i\in\{1,2,3,4\}$, labelled as in Fig. \ref{fig_models}. We label the two $\chi^i$'s on the same edge with the same number, so the number can be regarded as assigned to the edge. The four types of edges are denoted by $\mu(e)=1,2,3,4$. Besides, there are another two Majorana fermions $\eta^1, \eta^2$ on each site. The representation we choose is as follows
\begin{equation}
\begin{array}{lll}
\Phi^{12}=\sups{X}{1}; & \Phi^{23}=\sups{Z}{1}; & \Phi^{13}=\sups{Y}{1}; \\
\Phi^{41}=\sups{Z}{1}\sups{Y}{2}; & \Phi^{42}=-\sups{Y}{1}\sups{Y}{2}; & \Phi^{43}=\sups{X}{1}\sups{Y}{2}; \\
\end{array}
\end{equation}
Representation for $\Lambda^{ij}=i\eta^i\chi^j$ is chosen as
\begin{equation}
\begin{array}{ll}
\Lambda^{11}=\sups{Z}{1}\sups{X}{2}; &\Lambda^{12}=-\sups{Y}{1}\sups{X}{2};\\
\Lambda^{13}=\sups{X}{1}\sups{X}{2}; &\Lambda^{14}=\sups{Z}{2}; \\
\Lambda^{21}=\sups{Z}{1}\sups{Z}{2}; &\Lambda^{22}=-\sups{Y}{1}\sups{Z}{2};\\
\Lambda^{23}=\sups{X}{1}\sups{Z}{2}; &\Lambda^{24}=\sups{X}{2}. \\
\end{array}
\end{equation}

We use $\mu(e), s(e), t(e)$ to denote edge type, red site and blue site of an edge $e$, with mapping arrows from red sites to blue sites. With such a qubit representation, a Hamiltonian $H=\sum_{e} J_{\mu(e)} (i\ga^1_{s(e)}\ga^1_{(t)}+i\ga^2_{s(e)}\ga^2_{t(e)})$ on fermionic side is mapped to
\begin{equation}\begin{split}
H&=-\sum_{e}  J_{\mu(e)} \Lambda^{1\mu(e)}_{s(e)}\Lambda^{1\mu(e)}_{t(e)}\\
&=-\sum_{e} J_{\mu(e)} \sigma^{\mu(e)}_{s(e)} \sigma^{\mu(e)}_{t(e)} (\sups{X}{2}_{s(e)}\sups{X}{2}_{t(e)}+\sups{Z}{2}_{s(e)}\sups{Z}{2}_{t(e)}),
\end{split}
\end{equation}
which is Ryu's diamond model \cite{PhysRevB.79.075124}. This is an example of 3D quantum spin liquid model. Both Hamiltonians of Ryu's diamond model and Kitaev's honeycomb model pick up the unique ground state satisfying the plaquette constraints so there is no need to include constraints in the Hamiltonians. These plaquette terms appear as effective theories automatically in strong coupling limit. In fact, the same Hamiltonian can be constructed for a square lattice with two types of sites interspersed. Interested readers are referred to Ref.\ \onlinecite{PhysRevB.85.155119}.

\bibliography{references}

\end{document}